\documentclass[usenatbib]{mn2e}
\usepackage{graphicx}
\usepackage{amsmath}

\def\CN2{\mbox{$C_N^2$}}
\def\CT2{\mbox{$C_T^2 \ $}}

\def\sigmal2{\mbox{$\sigma ^{2}_{I} \ $}}

\title[On the comparison between MASS and G-SCIDAR techniques]{On the comparison between MASS and G-SCIDAR techniques}
\author[E. Masciadri et al.]{\vspace{1.7mm}
{E. Masciadri$^{1}$\thanks{E-mail: masciadri@arcetri.astro.it}, 
G. Lombardi$^{2}$, F. Lascaux$^{1}$}\\
$^{1}$INAF - Arcetri Astrophysical Observatory, Largo E. Fermi 5, I-50125 Firenze, Italy\\
$^{2}$European Southern Observatory, Casilla 19001, Santiago 19, Chile\\
}

\begin{document}
\date{Accepted 2013 November 7. Received 2013 November 7; in original form 2013 October 7}


\maketitle

\label{firstpage}
\setcounter {figure} {0}
\begin{abstract}
The Multi Aperture Scintillation Sensor (MASS) and the Generalized-Scintillation Detection and Ranging (Generalized SCIDAR) are two instruments conceived to measure optical turbulence (OT) vertical distribution on the whole troposphere and low stratosphere ($\sim$ 20~km) widely used in the astronomical context. In this paper we perform a detailed analysis/comparison of measurements provided by the two instruments and taken during the extended site testing campaign carried out on 2007 at Cerro Paranal and promoted by the European Southern Observatory (ESO). The main and final goal of the study is to provide a detailed estimation of the measurements reliability i.e dispersion of turbulence measurements done by the two instruments at different heights above the ground. This information is directly related to our ability in estimating the absolute value of the turbulence stratification. To better analyze the uncertainties between the MASS and the GS we took advantage of the availability of measurements taken during the same campaign by a third independent instrument (DIMM - Differential Imaging Motion Monitor) measuring the integrated turbulence extended on the whole 20~km. Such a cross-check comparison permitted us to define the reliability of the instruments and their measurements, their limits and the contexts in which their use can present some risk. \end{abstract}


\section{Introduction}

The optical turbulence vertical profilers (OTVP) are widely used in ground-based astronomy to estimate the stratification of the optical turbulence ($\CN2$ profiles) from the ground up to the top of the atmosphere ($\sim$ 20~km) or for sub-ranges of this interval \citep{roddier1981,rocca1974,vernin1983,avila1997,klueckers1998,fuensalida2004,egner2007,kornilov2003,wilson2002,azouit2005,avila2009,egner_masciadri2007,osborn2010,beckers2001,hickson2004,tokovinin2010}. Optical turbulence (OT) degrades the quality of images obtained with ground-based telescopes by introducing perturbations on the perfect wavefronts coming from the observed objects located outside the atmosphere at infinite distance. To correct the wavefront perturbations, recover the intrinsic angular resolution of telescopes and obtain images at diffraction limit, different typologies of adaptive optics (AO) techniques have been developed in the last decades \citep{beckers1993,davies2012}. However, all AO techniques depend on the OT status in the whole atmosphere. The knowledge of the spatio-temporal distribution of the OT in an astronomical site is therefore fundamental for many reasons: to optimize the AO techniques efficiency; to search for sites with favorable turbulence characteristics; to characterize the atmosphere of the existent observatories to prove the reliability of new AO techniques such as the wide-field-of-view AO \citep{neichel2009,vidal2010,basden2013}; to forecast the OT to schedule scientific programs to be carried out and instruments to be placed at the focus of telescopes. Different OTVPs have been developed in the last decades, each one characterized by some specific features.

The Generalized SCIDAR (hereafter GS: \citet{fuchs1994,fuchs1998}) is a well known technique based on the autocorrelation of scintillation maps produced by binary stars on the pupil of a telescope. This technique permits to reconstruct the turbulence stratification from the ground up to 20-25~km with a variable vertical resolution that scales as \citep{vernin1983}:

\begin{equation}
\Delta H(h) = \frac{{0.78 \cdot \sqrt {\lambda | h - h_{gs}|} }} {\theta }  
\label{res}
\end{equation}
\noindent
where $\lambda$ is the wavelength, $h$ is the height from the ground, $h$$_{gs}$ is the conjugated height under-ground (see Section \ref{gs}) and $\theta$ is the binary separation. Considering the typical values of the binaries separation and $h$$_{gs}$ used for the GS,  we can say that the GS vertical resolution is typically of the order of 1~km on the whole atmosphere. The GS is based on a solid and simple optical principle so that several GSs have been built by different teams worldwide in the last two decades and used for site characterization on many among the major astronomical Observatories in the world \citep{avila1997,avila2004,fuensalida2004,klueckers1998,mckenna2003,garcia2006,egner2007,egner_masciadri2007,fuensalida2008,masciadri2010,garcia2011a,garcia2011b,avila2011,masciadri2012}. The GS is, at present, one of the most reliable OTVP to be used as a reference for testing new OTVPs on the whole 20~km. The main limitation is that it requires a telescope of at least 1~m (preferably 1.5~m) pupil size and the brightness of the binary stars has to be $\le$ 5-6 mag.  This reduces the number of usable stars on sky. It is therefore not a good candidate to be used as an automatic monitor to be placed in astronomical observatories for systematic optical turbulence monitoring and/or for site searches but it is more suitable for dedicated experiments. 

The Multi Aperture Scintillation Monitor (MASS) entered in the astronomic context in more recent years \citep{kornilov2003}. It is based on a principle proposed by Ochs in 1976 \citep{ochs1976} that implies the observation of scintillation maps produced by single stars on the telescope's pupil. The turbulence profile is reconstructed thanks to the analysis of the spatial structure of the scintillation map through a set of different spatial filters and the use of a set of weighting functions. Kornilov introduced, as spatial filters, a set of four small concentric and circular apertures selected on the telescope's pupil \citep{kornilov2003}. The main attractive features of the MASS are: (a) a better sky coverage than the GS (single stars are much more numerous than binary stars with magnitude $\le$ 5-6~mag) and (b) a small pupil of the telescope (of the order of 8-15~cm) well suitable therefore to be employed as an automatic monitor. The main drawbacks are: (a) a lower vertical resolution that scales as $\Delta h$ $\sim$ 0.5$\cdot$h \citep{tokovinin2003a}. More precisely it provides OT profiles sampled on 6 layers centered at 0.5~km, 1~km, 2km, 4~km, 8~km and 16km a.g.l.; (b) the MASS is insensitive to the optical turbulence near the ground (below $\sim$ 0.5~km). It is therefore commonly considered as a free-atmosphere turbulence monitor. Turbulence ground layer contribution has been frequently estimated by subtracting the MASS estimate from the Differential Image Motion Monitor (DIMM)\footnote{The latter provides estimates of the integrated turbulence on the whole atmosphere (along the optical path).};  (c) because of the particular weighting functions (triangle shape) of the MASS, this method does not permit to identify precisely the height of the boundary layer and in general the height separating a layer from the contiguous one. Turbulence developed inside a vertical slab [h$_{1}$,h$_{2}$]~km is distributed indeed in two different contiguous weighting functions. Even if this is not particularly critical for the site selection it can create problems in some other contexts and applications such as the AO. MASS and DIMM have been widely used in the last decade to characterize the astroclimatic parameters of many astronomical sites in the world \citep{tokovinin2003b,lawrence2004,kornilov2010,lombardi2010,dali2010,thomas-osip2012}) but it has been developed by just one scientific group. An intense use of this instrument has been done also for new generation telescopes site searches such as those aiming at the identification of the Thirty Meters Telescope (TMT) \citep{schoeck2009} and the European Extremely Large Telescope (E-ELT) \citep{vazquez-ramio2012} sites. 

Curiously, little has been done so far in terms of analysis of the MASS reliability with other OTVP. The unique study has been carried out so far by Tokovinin \citep{tokovinin2005} in which MASS and GS measurements taken simultaneously at Mauna Kea have been statistically compared on a sample of four nights\footnote{Most of the analysis on reliability has been done so far with simulations and post verification of the exact reconstruction of the vertical stratification \citep{tokovinin2003a}. }. The conclusions of that paper state that a very good agreement (better than 20$\%$) of the integral of the turbulence in the free atmosphere as measured by the two instruments is observed. The authors report that a very satisfactory agreement is reached for layers at 8 and 16~km while the MASS systematically overestimates the OT in the layer at 0.5~km and underestimates the OT in layers at 2 and 4~km. These errors have been ascribed to the restoration technique. In other words, the total integral of the six layers is well reconstructed by the MASS but its distribution among the six predefined layers presents some systematic biases.

Because measurements from a more recent and statistical rich site testing campaign done at Cerro Paranal (and called PAR2007) are now available \citep{dali2010}, we have performed in this paper a detailed cross-comparison of MASS and GS measurements with the aim to provide a quantitative estimation of the accuracy of the turbulence stratification and an as precise as possible analysis of the reliability of the two instruments. This observed OT stratification represents, indeed, for us a reference to calibrate a mesoscale atmospherical model to be used to predict the OT stratification \citep{masciadri2013a}. It is therefore for us extremely important to quantify the absolute values of these quantities. 

In Section \ref{obs} we present the PAR2007 site testing campaign instruments and measurements that we treated. In Section \ref{instr} we briefly remind the main physical principles which the two techniques of the GS and MASS are based on with some basic information on the DIMM too. In Section \ref{analy} we present the comparison of the different astroclimatic parameters (seeing in the free atmosphere, seeing in each individual layer, isoplanatic angle and wavefront coherence time) provided by the two instruments with the associated discussion. In Section \ref{disc} we comment our results with respect to the literature. In Section \ref{concl} we present the conclusions of our study.

\section{Observations}
\label{obs}

\begin{figure}
\centering
\includegraphics[width=5.5cm]{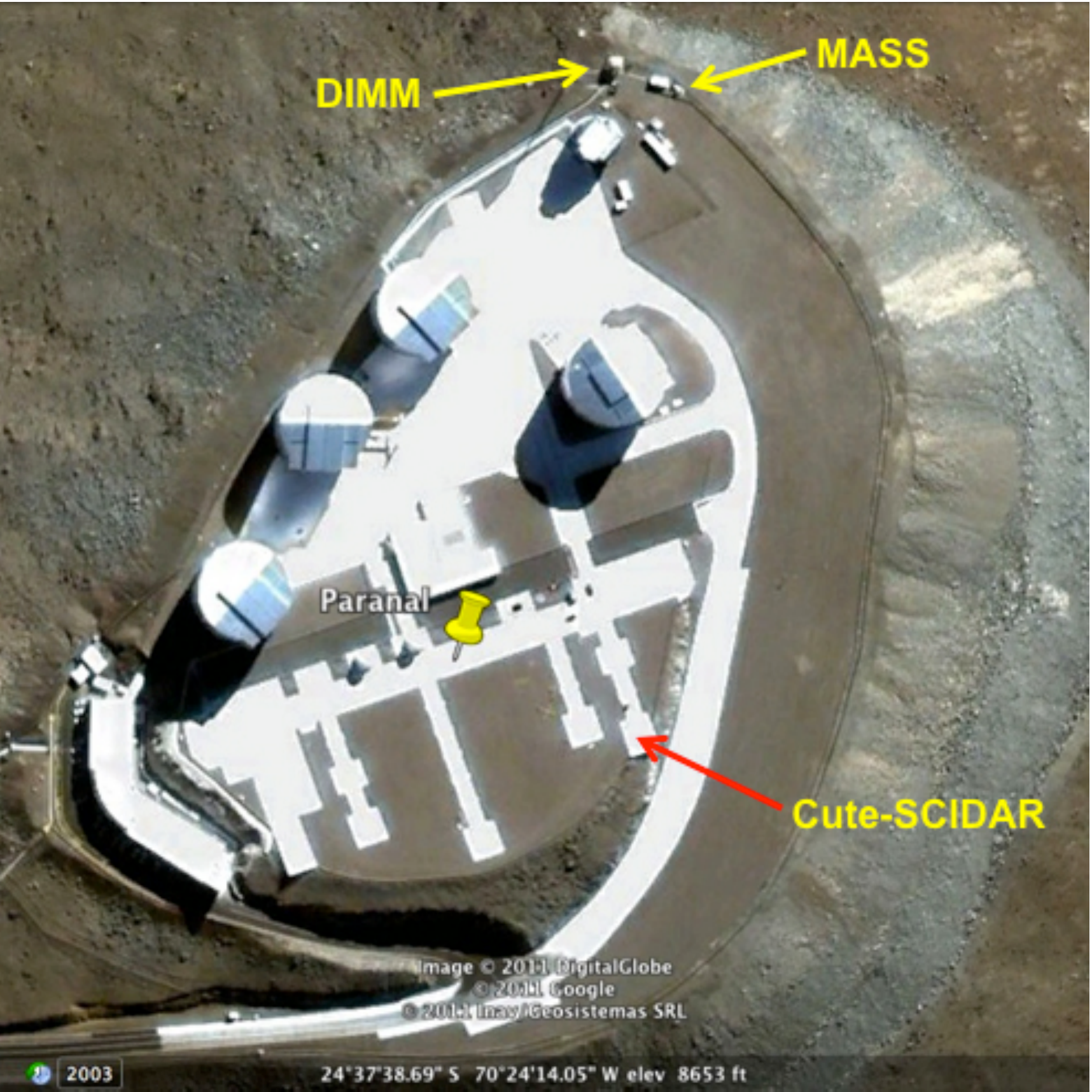}
\caption{Satellite image of the Paranal Observatory from Google Earth. Yellow arrows indicate the position of DIMM, MASS and the Generalized SCIDAR. The distance between the DIMM and the Cute-SCIDAR is 205~m. The distance between the MASS and the DIMM is 20~m. The telescope UT4 is located at mid-path between the DIMM and the Cute-SCIDAR. The Cute-SCIDAR is a Generalized SCIDAR.
\label{map}} 
\end{figure}

Measurements we treated in this paper belong to the PAR2007 site testing campaign that took place at Cerro Paranal in November-December 2007 \citep{dali2010} and was promoted by ESO in the context of the E-ELT Design Study. A set of instruments, among those a MASS, a GS and a DIMM have been run during 20 nights in this couple of months. Simultaneous measurements of GS and DIMM on 20 nights and simultaneous measurements of GS and MASS on a sub-sample of 14 nights are available (Table \ref{tab:nights}). In 2 nights over 14 there are just a few simultaneous measurements. We considered therefore a sub-sample of 12 nights. Figure \ref{map} shows the position of the three instruments on the summit of Cerro Paranal. The GS is placed at the focus of an Auxiliary Telescope (AT). The GS and DIMM are located at 5 and 6 meters above the ground respectively. The distance between the two instruments is roughly 205~m over a basically flat surface.

\begin{table}
\caption{List of nights in which simultaneous DIMM and GS measurements are available in the PAR2007 site testing campaign (20 nights). The asterisk indicates the sub-sample of nights in which MASS and GS simultaneous measurements are available (14 nights). The double asterisk indicates the nights in which just a few (respectively 1 and 2) seeing simultaneous measurements are available. These two nights are not used in the statistic analysis.\label{tab:nights}}
\begin{center}
\begin{tabular}{c c c}
\hline
\multicolumn{3}{c}{Observation nights of PAR2007 Site Testing Campaign} \\
\multicolumn{3}{c}{with DIMM, GS and MASS} \\
\hline
2007/11/10  & 2007/11/21$^{*}$ & 2007/12/19$^{*}$ \\
2007/11/11  & 2007/11/22$^{*}$ & 2007/12/20$^{*}$ \\
2007/11/13  & 2007/11/24 & 2007/12/21$^{*}$\\
2007/11/14  & 2007/12/15$^{**}$ & 2007/12/22$^{*}$\\
2007/11/17  & 2007/12/16$^{**}$ & 2007/12/23$^{*}$ \\
2007/11/18$^{*}$ & 2007/12/17$^{*}$ & 2007/12/24$^{*}$\\
2007/11/20$^{*}$ & 2007/12/18$^{*}$ & \\
\hline
\end{tabular}
\end{center}
\end{table}

\section{Instruments}
\label{instr}
\subsection{Generalized SCIDAR: Cute-SCIDAR}
\label{gs}

The SCIDAR technique (Scintillation Detection and Ranging) has been originally proposed by \citet{rocca1974}  and \citet{vernin1983} to measure the vertical optical turbulence distribution (the refractive index structure constant $C_{N}^{2}$ profiles) in the troposphere. The technique relies on the analysis of the scintillation maps generated by binary stars on the pupil plane of a telescope. The standard SCIDAR technique (called Classic Scidar) is insensitive to the turbulence near the ground. This fact represented in the past an important limitation for monitoring turbulence for astronomical applications because it is known that most of the turbulence develops in the low part of the atmosphere. To overcome this limitation \citet{fuchs1994} and \citet{fuchs1998} proposed a generalized version of the SCIDAR (called Generalized SCIDAR) in which the detector is virtually conjugated below the ground at a distance $h_{gs}$ permitting to extend the measurement range to the whole atmosphere (from the ground up to $\sim$ 20-25 km). 

The GS is based on the observation of binaries having an angular separation typically of $\sim$ 3-10 arcsec, magnitude $\le$ 5-6 mag and a $\Delta$$m$ $\sim$ 1 mag. When two plane wavefronts coming from a binary and propagating through the atmosphere meet a turbulent layer located at a height $h$ from the ground, they produce, on the detector plan, optically placed below the ground at a few kilometers ($h_{gs}$), two scintillation maps characterized by typical shadows appearing in couple at a distance $d$. Such a distance is geometrically related to the position of the turbulent layer as $d$ = $\theta$($h$+$h_{gs}$). The calculation of the auto-covariance (AC) of the scintillation map, normalized by the autocorrelation of the mean image, produces the so called 'triplet' i.e. three peaks. The central peak of the triplet is located at the centre of the AC frame; the lateral peaks are symmetrically located at a distance $d$ from the centre. 
The amplitude of the lateral peaks is proportional to the auto-covariance of the scintillation map therefore to the strength of the turbulent layer.
In summary the amplitude of the later peaks of the triplet and their distance from the central peak give us the information on the strength and the height h of the turbulent layer respectively. In a multi-layers atmosphere, different turbulent layers located at different height $h_{i}$ produce several triplets all centred in the centre of the AC frame but with lateral peaks located at different distances $d_{i}$ from the centre of the AC frame. The $\CN2$ profiles are obtained taking into account the information provided by every turbulent layer all together. From an analytical point of view, as it is indicated by \citet{avila1997}-Eq.1, to obtain the $\CN2$ profiles one has first to calculate the difference of the perpendicular and parallel sections of the normalized auto-covariance (AC), and then inverting this equation (known as Fredholm equation). 

The GS used in the PAR2007 is called CUTE-Scidar III (see Fig.\ref{map}). It has been developed by the Instituto de Astrof\'isica de Canarias (IAC) team (V\'azquez-Rami\'o et al., 2008). CUTE-Scidar III instrument provides $\CN2$ profiles in real time without dome and mirror seeing components and it has been used since 2002 to extensively monitor the optical turbulence at El Roque de Los Muchachos and Teide observatories in Canary Islands \citep{fuensalida2004,garcia2011a,garcia2011b}. The dome and mirror seeing is quantified using a technique based on evenness properties with the Fourier analysis \citep{fuensalida2008}. The GS principle, indeed, permits to measure $\CN2$ profiles including the turbulence contribution coming from the dome. A dedicated procedure is necessary to disentangle that from the turbulence provided by the atmosphere. The temporal sampling of the raw measurements is $\sim$ 1 minute. 
$\CN2$ profiles are sampled along the z-axis at 300~m but this spatial scale is totally arbitrary. 
The intrinsic $\CN2$ profiles spatial resolution is given by Eq.\ref{res}. Measurements of the PAR2007 campaign done with the CUTE-Scidar III have been corrected \citep{masciadri2012} to eliminate the error introduced by the normalization of the autocorrelation of the scintillation maps by the autocorrelation of the mean pupil \citep{johnston2002,avila2009}. 
This error, if it is not corrected, induces an overestimation of the $\CN2$(h) that depends on many parameters related to the optical set-up and the observed binaries. More precisely, it depends on the diameter of the pupil of the telescope $D$, the height h$_{gs}$ at which the detection plane is conjugated below the ground, the difference $\Delta$m between the stellar magnitudes of the binaries, the angular separation of the binary $\theta$ and the ratio $e$ between the central obscuration D$^{*}$ and the telescope pupil (e =
  D$^{*}$/D).

\subsection{MASS} 
\label{mass}

\begin{table}
\caption{Configuration of the MASS weighting functions extremes: triangle's bases and peaks. Units in meters (m).\label{table1}}
\begin{center}    
\begin{tabular}{c c c c c}
\hline
Layer & $h$ & $min_{i}$ & $max_{i}$\\
\hline
1 & 500   &  250  & 1000\\
2 & 1000  &  500  & 2000\\
3 & 2000  &  1000 & 4000\\
4 & 4000  &  2000 & 8000\\
5 & 8000  &  4000 & 16000\\
6 & 16000 &  8000 & - \\
\hline
\end{tabular}
\end{center}
\end{table}

\begin{figure}
\centering
\includegraphics[width=7cm]{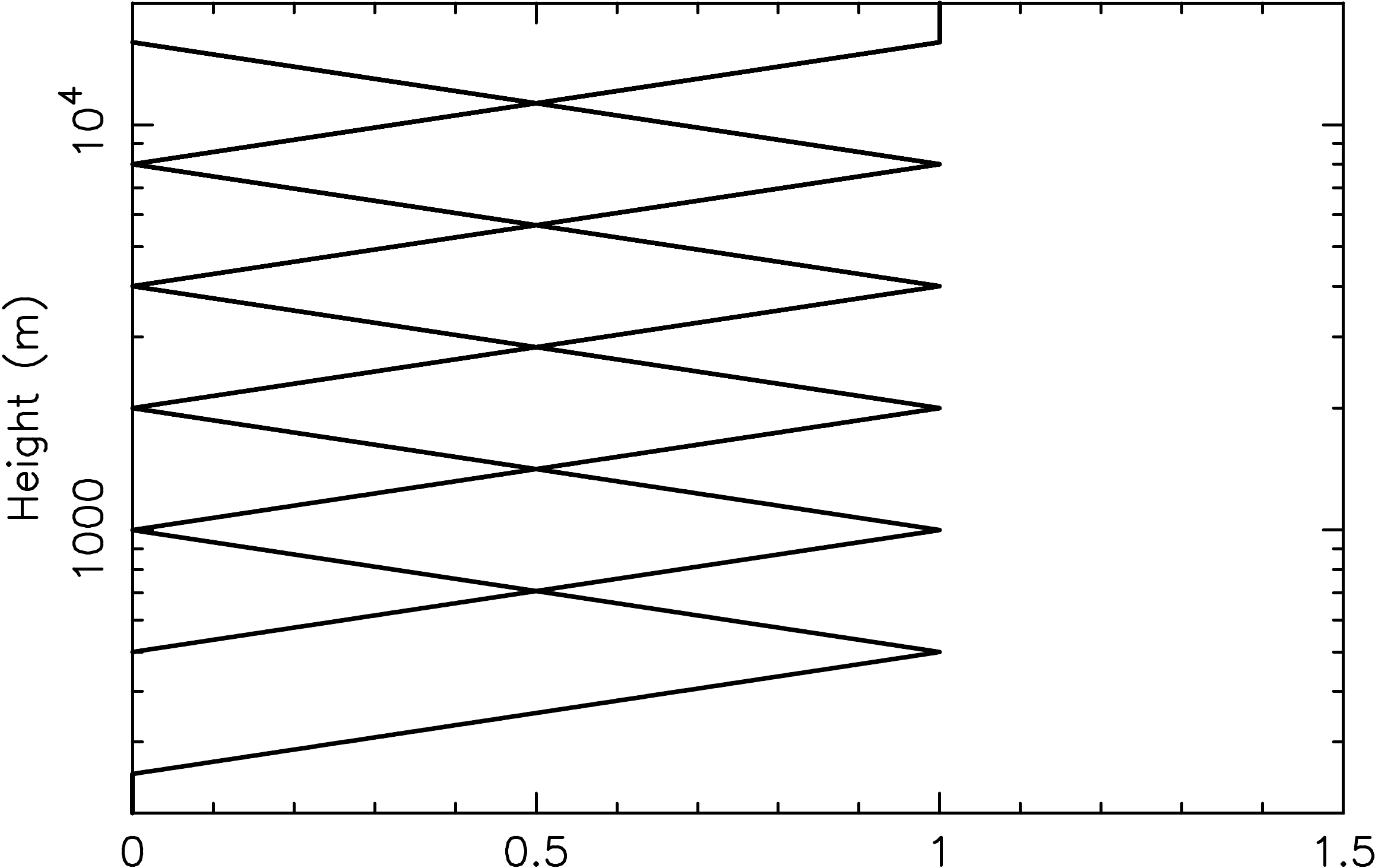}
\caption{MASS weighting functions as a function of the height from the ground and valid for the instrument 6-layers configuration. Y-axis is in logarithmic scale. 
\label{wf}} 
\end{figure}

As anticipated in the Introduction, the MASS has been introduced in the astronomical context by Kornilov \citep{kornilov2003} with a slightly modified version of a technique proposed by \citet{ochs1976}. The MASS reconstructs the turbulence stratification in six layers distributed in the 20~km above the ground starting from the analysis of scintillation indices (four normal and six differential indices) of scintillation maps produced by single stars on a set of four small concentric and circular apertures selected on the telescope's pupil. Scintillation is measured by photo-multipliers. The vertical stratification is obtained by fitting a set of measured scintillation indices with a model having a small and fixed number of turbulent layers. The layers are located at 6 heights: 0.5, 1, 2, 4, 8 and 16~km above the ground. Their positions respect therefore the following law:\newline
\begin{equation}
h_{i}=2h_{i-1}
\end{equation}
where $i$ =1,..,6 and $h_{1}$ = 500~m.
The scintillation indices depend on the integral of the $\CN2$(h), the turbulent spectrum, the Fresnel diffraction term and the pupil filter. The product of the turbulent spectrum times the Fresnel diffraction term times the pupil filter constitutes the weighting function $W(h)$. The integral of the turbulence in each layer $J_{i}$ (with i=1,..,6 ) is the integral of the $\CN2$ times the weighting function $W(h)$ (see Eq.1 in \citet{tokovinin2003b}):\newline
\begin{equation}
J_{i}=\int_{min_{i}}^{max_{i}}C_{N}^{2}(h)\cdot W(h)dh
\end{equation}
where $J$ is expressed in (m$^{1/3}$) and $min_{i}$ and $max_{i}$ represent the extremes of the layer $i$. The MASS weighting functions (WFs) are a sequence of triangles whose peaks are located at the layer heights (0.5, 1, 2, 4, 8 and 16~km) and the base of the triangles are reported in Table \ref{table1}. The sequence of the weighting functions in logarithmic scale along the whole atmosphere is shown in Fig.\ref{wf}. In this figure it is visible how each triangle is interlaced with the previous and successive one. For each height h in the [0.5, 16]~km range is valid the condition that the value of the weighting function $W(h)$ is always equal to 1. This condition guarantees the conservation of the turbulence energy in the vertical slab [0.5, 16]~km. The MASS device used during the PAR2007 site testing campaign was developed by the Kornilov's group at the Sternberg Astronomical Institute in Russia to be employed in the E-ELT site testing campaigns in the northern part of Chile and in Argentina. The temporal sampling of the raw measurements is of around 1 minute. The MASS position at the summit of Cerro Paranal is shown in Fig. \ref{map}. 

\subsection{DIMM} 
\label{dimm}

DIMM measurements are routinely done at Cerro Paranal since 1988 in coincidence of the Very Large Telescope Site Testing campaign. The DIMM measures the integrated turbulence energy in the whole atmosphere (the whole optical path covered by the wavefront). The DIMM principle is based on the measurement of the variance of the fluctuations of the angle of arrival of the wavefront coming from single stars and passing through two small holes separated by a certain distance d within the pupil of a telescope \citep{sarazin1990}. The telescope pupil size is typically of 35-40~cm, the holes size $D$ = 6~cm and the distance $d$ = 14~cm. Values of $D$, $d$ and the pupil size of the small telescope can change depending on the prototype provided $d$ $\ge$ 2$\cdot$D. The seeing $\varepsilon$ is retrieved from the longitudinal $\sigma^{2}_{l}$ and transversal $\sigma^{2}_{t}$ variances of the fluctuations of the angles of arrival. The temporal sampling of measurements is of around 1 minute. In this paper DIMM measurements will be compared to the GS ones in order to have more solid arguments for our conclusions on GS versus MASS seeing comparison. We note that the methods used by the DIMM to calculate $\theta_{0}$ \citep{krause-polstorff1993} and $\tau_{0}$ \citep{sarazin2002} are approximated methods. We will not use therefore DIMM estimations as a reference to support the comparison between the GS and MASS estimation of these two parameters. It has been already put in evidence in the literature that important relative errors with respect to the GS taken as a reference can be obtained for $\tau_{0}$ using such methods \citep{masciadri2006,masciadri2013c}. In \citet{masciadri2013c} preliminary results on a comparison GS vs. DIMM for $\theta_{0}$ have been presented. 
The DIMM location at the summit of Cerro Paranal is shown in Fig.\ref{map}. 

\section{Analysis}
\label{analy}

\subsection{Seeing}
\subsubsection{MASS versus GS}
\label{mass_gs_see}

The comparison MASS vs. GS has been done projecting the $\CN2$ profiles of the GS on the weighting functions (Fig.\ref{wf}) that means re-binning the $\CN2$ of the GS on the weighting functions of the MASS. This permits us to integrate the new GS $\CN2$ profiles in each triangle and therefore to compare the relative MASS and GS seeing (or J) in each layer as well as in the sum of all the six layers.
To perform the statistical analysis of the comparison we used the bias (Eq.\ref{eq:bias}), the root mean squared error (RMSE) (Eq.\ref{eq:rmse}) and the regression line passing by the origin\footnote{It has been proved in recent studies \citep{lascaux2013} that the correlation coefficient is not a suitable statistical estimator for this kind of analysis. }. The bias provides us information on systematic errors, the RMSE on the statistical errors plus the systematic errors obtained by the two instruments. The regression line tells us how close/far is the distribution of measurements from a $y$ = $x$ analytical linear relationship. 

\begin{equation}
BIAS = \sum_{i=1}^{N}\frac{Y_i-X_i}{N}
\label{eq:bias}
\end{equation}

\begin{equation}
RMSE = \sqrt{\sum_{i=1}^{N}\frac{(Y_i-X_i)^2}{N}}
\label{eq:rmse}
\end{equation}
where $X_{i}$ and $Y_{i}$ are the individual GS and the MASS seeing values respectively. $N$ is the total number of elements in the sample. From the bias and the RMSE it is possible to retrieve the bias-corrected RMSE:\newline
\begin{equation}\begin{split}
\sigma=\sqrt{\sum_{i=1}^{N}\frac{[(X_i-Y_i)-(\overline{X_i-Y_i})]^2}{N}}\\ =\sqrt{RMSE^2-BIAS^2}
\label{eq:sigma}
\end{split}\end{equation}

$\sigma$ contains only the statistical errors and it will be discussed in some cases.

\begin{figure*}
\begin{center}
\includegraphics[width=15cm]{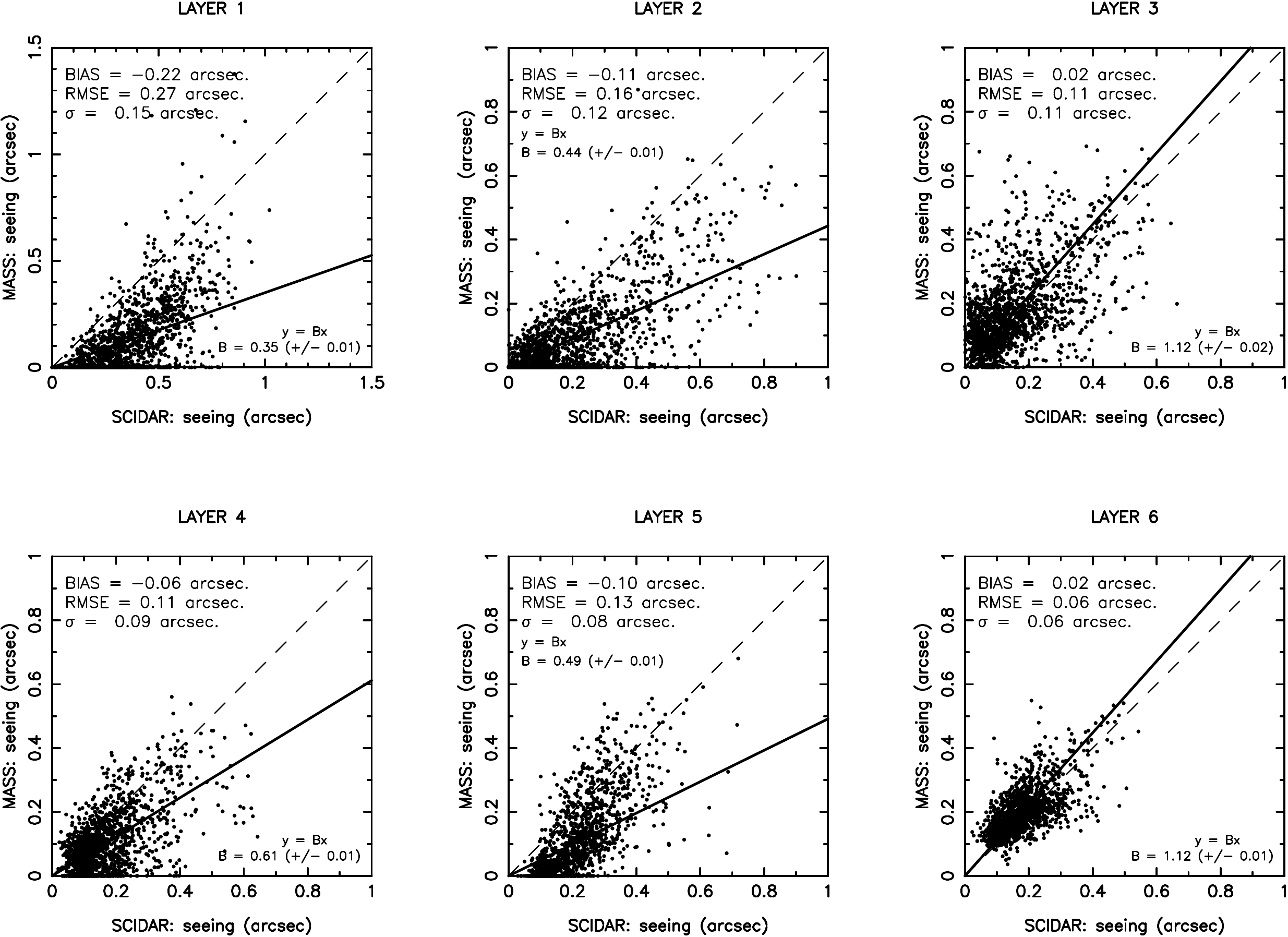}
\end{center}
\caption{Scattering plots of the MASS and GS measurements of turbulence (seeing values) related to the nights in which 
simultaneous measurements obtained by the two instruments are available: 12 nights (see Table \ref{tab:nights}). Measurements are 
sampled on a timescale of 1-minute. GS measurements are re-binned as if they were weighted by the triangle MASS weighting function in order to 
be compared to MASS measurements. Each figure of the panel reports the measurements distribution related to one of the six layers. The bias refers to ${\varepsilon}_{MASS} - {\varepsilon}_{SCIDAR}$. \label{fig1}} 
\end{figure*}

\begin{figure*}
\centering
\includegraphics[width=6cm]{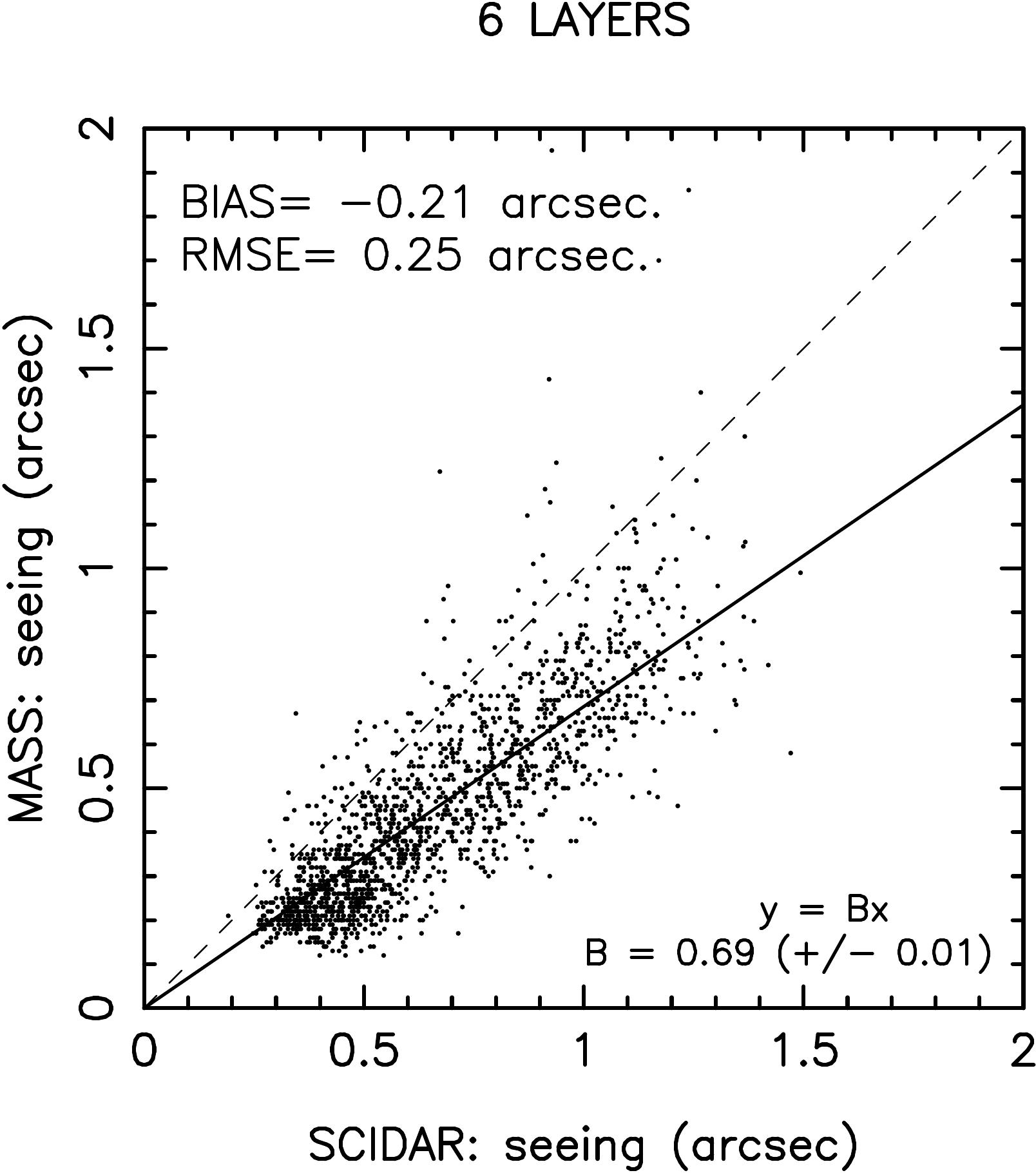}
\includegraphics[width=6cm]{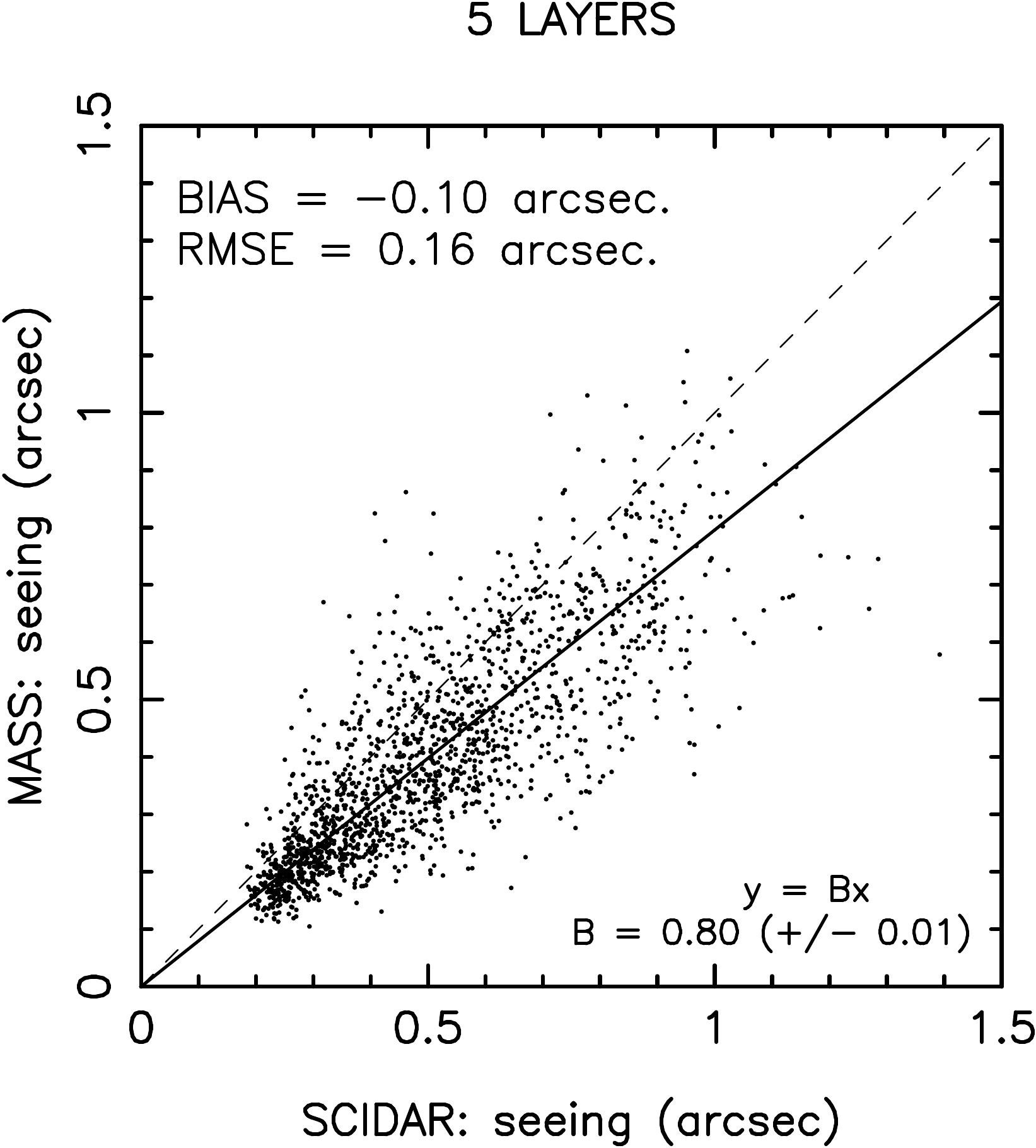}
\caption{As Fig.\ref{fig1}. Scattering plot of the seeing as measured simultaneously  by the MASS and the GS on the sub-sample of 12 nights (Table \ref{tab:nights}). Left: measurements coming from the first (layer 1) to the sixth (layer 6) layer are considered. Right: measurements coming from the second (layer 2) to the sixth (layer 6) layer are considered.
\label{fig2}} 
\end{figure*}

\begin{table}
\caption{Number of measurements available during each night from the GS and the MASS to be used for the statistical analysis. \label{table_meas}}
\begin{center}    
\begin{tabular}{c c c }
\hline
Night &  N. of GS meas. & N. of MASS meas. \\
\hline
18/11/2007 & 391  &   64 \\
20/11/2007 & 371  &   103 \\
21/11/2007 & 147  &   57 \\
22/11/2007 & 320  &  203 \\
17/12/2007 & 345  &  33 \\
18/12/2007 & 454  &  64\\
19/12/2007 & 440  &  184\\
20/12/2007 & 483  &  364 \\
21/12/2007 & 492  &  97 \\           
22/12/2007 & 491  &  302 \\
23/12/2007 & 402  &  270 \\
24/12/2007 & 452  &  166 \\
\hline
\end{tabular}
\end{center}
\end{table}

 \begin{table*}
  \centering
  \caption{Averages, standard deviations and relative errors of the seeing as measured simultaneously by the GS and the MASS on the sub-sample of 12 nights (Table \ref{tab:nights}). Measurements are resampled on a time scale of 1 minutes. Relative error is defined as (Eq.\ref{eq:rel}). Columns 8 and 9 report the relative errors expressed in terms of J (see Eq.\ref{j_epsi}) in this paper (col. 8) and in paper \citep{tokovinin2005} (col. 9).\label{tab:stat}
}
   \begin{tabular}{|c|c|c|c|c|c|c|c|c|}
  \hline
  \multicolumn{2}{|c|}{ } & \multicolumn{2}{|c|}{SCIDAR} & \multicolumn{2}{|c|}{MASS}  &  \multicolumn{3}{|c|}{ } \\
  \hline
         Height      &     & Average  & std.             & Average & std. &   Relative Error  & Relative Error (J) & Relative Error (J) -T05 \\ 
           (km)     &    &  (arcsec) &  (arcsec)              & (arcsec) &  (arcsec) &  ($\%$)  &  ($\%$)  &  ($\%$)  \\                                
  \hline                
  & 6 layers  &   0.66     &  0.26     &  0.45       & 0.23  & -32  & -48 & -5 \\
  & 5 layers  &    0.50    & 0.21      &   0.40      &  0.19 & -20 & -30 & -\\
  \hline
  0.5 & Layer 1   &    0.34   &   0.18     &   0.12      &  0.17 & -65 & -82&  +100 \\
  1& Layer 2   &    0.20   &   0.17     &     0.09    & 0.12  & -55 & -71& -41 \\ 
  2& Layer 3   &   0.15    &    0.12    &   0.17       & 0.13 & +13 & +25& -67\\
  4& Layer 4   &   0.16    &  0.10      &    0.10      & 0.09  & -37 &-60 & -57\\
 8&  Layer 5   &   0.20    &   0.10     &     0.10    & 0.12 & -50 & -71& +14\\
  16 &Layer 6   &    0.17   &     0.08   &       0.19  & 0.07 & +18 & +20 & -25\\
  \hline
  \end{tabular}
 \end{table*} 
 
Figure \ref{fig1} shows the scattering plots of MASS and GS measurements of turbulence (seeing values) in each of the six layers and related to the nights in which simultaneous measurements obtained by two instruments are available (12 nights - Table \ref{tab:nights}). Fig. \ref{fig2} shows the same scattering plot for the seeing as measured by the MASS and the GS integrated on all the six layers (from the layer located at 0.5~km up to the layer located at 16~km) and on the five layers (from layer located at 1~km up to the layer located at 16~km)\footnote{We preferred to treat in these figures the seeing and not the integrated turbulence J (see later Eq.\ref{j_epsi}) because J should be represented in log-scale and the distribution of the points as well as the regression line can provide distorted visual effects in log-scale.}.  Table \ref{table_meas} reports the number of measurements from both GS and MASS used for this analysis. To perform the comparison on simultaneous measurements, the data-set has been resample on a time scale of 1 minute. We verified that results (Fig.\ref{fig1}, Fig.\ref{fig2} and Table \ref{tab:stat}) are substantially the same if we use a time scale of 10 minutes. Also we note that in Fig. \ref{fig1} there are frequent unrealistic zero values by the MASS. These features are produced by the restoration technique as said in Section 3 of \citet{tokovinin2005}. These values have not been considered in the statistical analysis.
MASS measurements have been treated with the data-reduction software  version Atmos 2.3 that means the same version used (for seeing and isoplanatic angle) in the analysis done in the context of the Thirty Meter Telescope (TMT) site testing study for the selection of the TMT site \citep{schoeck2009} and E-ELT site testing study for the selection of the E-ELT site \citep{vazquez-ramio2012}. Fig.\ref{fig1} tells us that, with exception of layer 3 and 6, the MASS presents in all the other layers, an underestimation of the seeing with respect to the GS (expressed by the bias). If we consider the total seeing on the whole six or five layers (Fig. \ref{fig2}) the underestimation of the MASS is still observed even if, in the five layers case, it is weaker. To better appreciate from a quantitative point of view these differences, Table \ref{tab:stat} reports averages, standard deviations and relative errors of the seeing as measured by the MASS and the GS in each layer and on the total six and five layers. The relative error is calculated as:\newline
\begin{equation}
r=\frac{\varepsilon _{MASS}-\varepsilon _{GS}}{\varepsilon _{GS}}\times100
\label{eq:rel}
\end{equation}
We can observe that, even if in absolute terms, all the differences are within 0.22 arcsec, the relative errors can be not necessarily negligible. Layers located at 0.5, 1 and 8~km (layers 1, 2 and 5) present the largest discrepancies between GS and MASS with relative errors respectively of -65, -55 and -50~$\%$. Layers located at 2 and 16~km (layers 3 and 6) show the smallest relative errors respectively of 13 and 18~$\%$. The relative error for the total integrated turbulence on the whole six layers is -32~$\%$. If we do not consider the layer at 0.5~km, i.e. the turbulence integrated in the five layers (from 2 to 6), the relative error remains within -20~$\%$. 
The same analysis could be done in terms of J i.e. the integral of the turbulence on the whole atmosphere that is related to the seeing $\varepsilon$ by a 5/3 power (see for example \citep{masciadri2010}):\newline
\begin{equation}
J = 9 \cdot 10^{-11} \cdot \lambda ^{1/3}  \cdot \varepsilon ^{5/3} 
\label{j_epsi}
\end{equation}
where $\lambda$ is the wavelength and J unit is [m$^{1/3}$]. Due the power 5/3, the relative errors in J are larger as can be seen in column 8 of Table \ref{tab:stat}\footnote{The calculation of the relative error in J terms permits to discuss our results with respect to the \citet{tokovinin2005} paper.}.

\begin{figure*}
\begin{center}
\includegraphics[width=6cm,angle=-90]{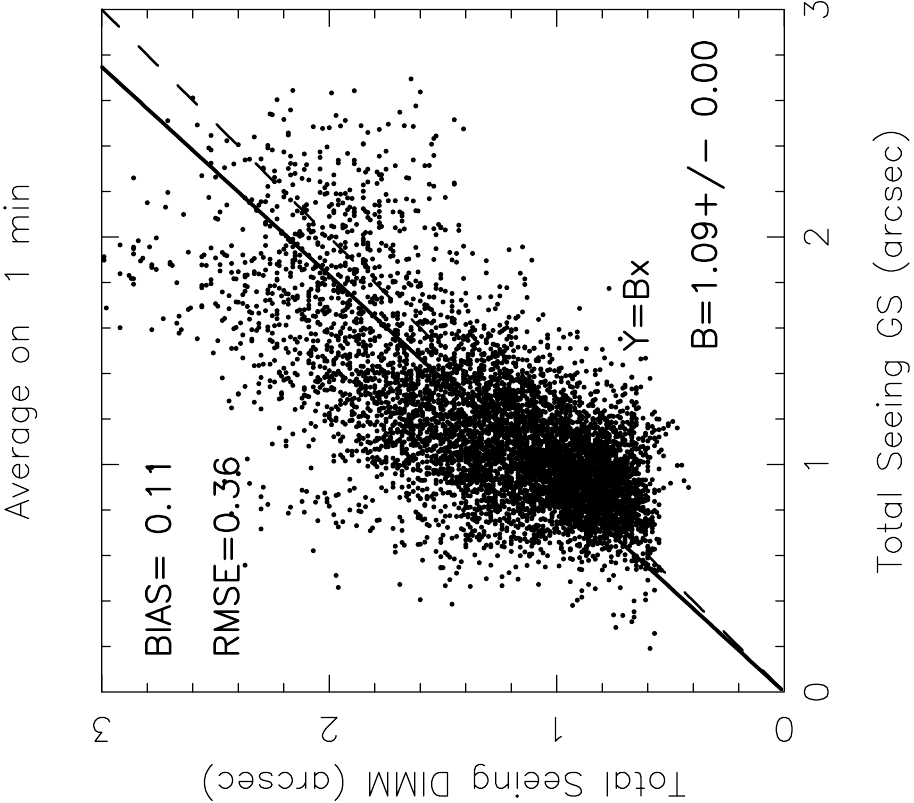}
\includegraphics[width=6cm,angle=-90]{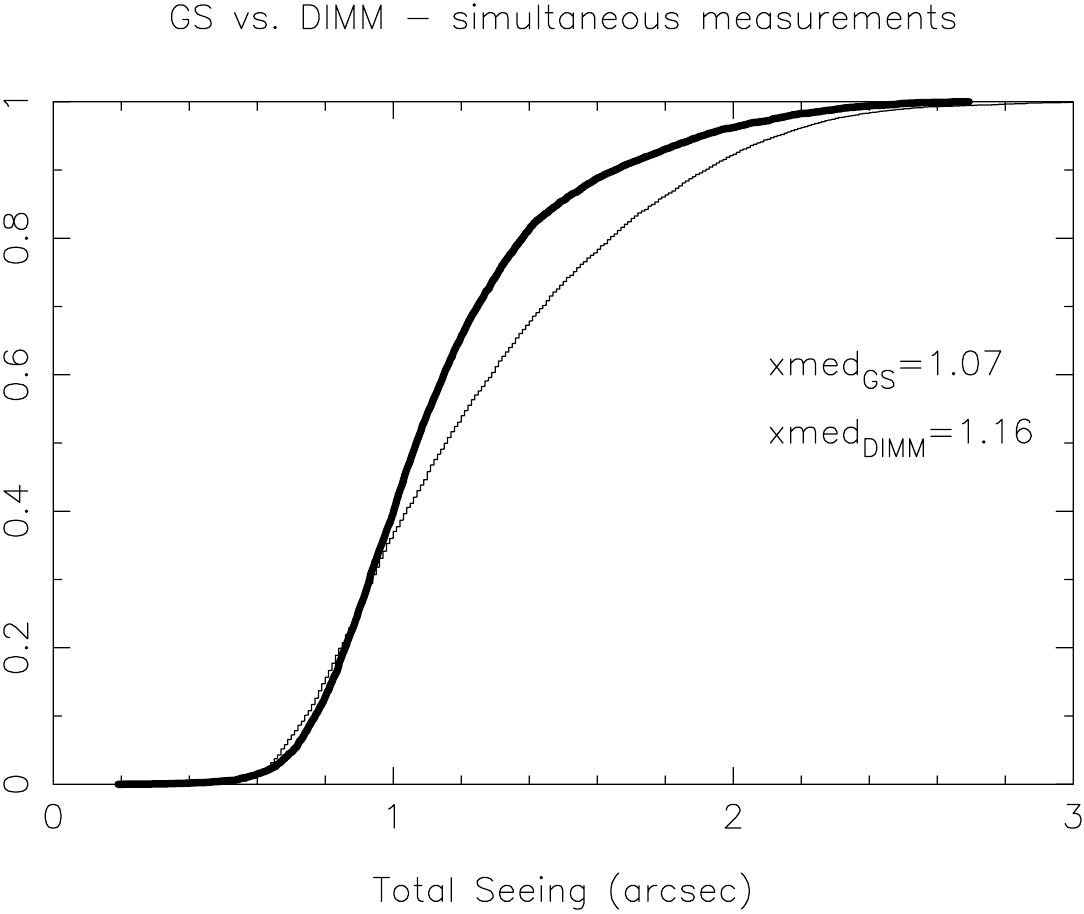}
\end{center}
\caption{Left:  Distribution of the DIMM and GS measurements of total turbulence (seeing values) related to the nights in which 
simultaneous measurements obtained by the two instruments are available: 20 nights (see Table \ref{tab:nights}). Measurements are 
sampled on a timescale of 1 minute. The bias refers to ${\varepsilon}_{DIMM} - {\varepsilon}_{SCIDAR}$. Right: cumulative distribution of the total seeing related to the whole atmosphere obtained by the GS and the DIMM. The median values obtained for the two instruments are reported in the figure.
\label{fig5}} 
\end{figure*}

To complete the analysis, Fig.\ref{fig4} in Annex shows the temporal evolution of the J values as measured by the MASS and the GS on the total six layers during the whole night of observation for all the 12 nights (Table \ref{tab:nights}). The temporal evolution provides information on the trend during the night. The quantitative estimation is done in any case by Table \ref{tab:stat}. It has been decided to represent the integrated turbulence $J$ (and not the seeing) just because, from a visual point of view, the dynamic between the measurements of the two instruments is better suited for a cross-comparison and our interest is to put in evidence if there are systematic biases. Fig.\ref{layer1} to Fig.\ref{layer6} show the temporal evolution of MASS and GS integrated $J$ in each individual layer for the sub-sample of 12 nights (Table \ref{tab:nights}) in which measurements are available. All the figures have been resampled on a time scale of 30 minutes to make the figures more readable. From these figures it is visible that the underestimation that has been quantified in Table \ref{tab:stat} in some layers is not the result of sporadic events but it is typical of most of the nights. This tells us that, even if the two instruments did not always look at the same direction, the underestimation can hardly be attributed to a different line of sight. On the other side, looking at Fig.\ref{layer1} to Fig.\ref{layer6} it is possible to observe the substantial good agreement of turbulence estimation in layers located at 2 and 16~km (layers 3 and 6). At the same time it is possible to note that the trend of the turbulence observed by the two instruments is substantially well reconstructed, even in those cases in which an underestimation is observed. We finally note that (see Fig.\ref{fig1}) the $\sigma$ in the individual layers is within [0.06-0.15] arcsec range with the largest $\sigma$ in the layer closest to the ground and the smallest $\sigma$ in the highest layer as it is expected. This tells us that the turbulence increases its horizontal homogeneity (small $\sigma$) when we go at high heights. The $\sigma$ increases near the ground due to the orographic effects.
 
To have some further elements that could confirm us that the problem in these estimates is represented by the MASS and not the GS we performed a comparison between GS and DIMM measurements that is discussed in the next Section.

\subsubsection{DIMM versus GS}

Fig. \ref{fig5}-left shows the scattering plot of the total turbulence as measured by the DIMM and GS on those nights of PAR2007 campaign in which simultaneous measurements are available (20 nights - see Table \ref{tab:nights}). Table \ref{table_meas_gsdimm} reports the number of measurements from both GS and DIMM used for this analysis. Measurements are resampled on a time scale of 1 minute\footnote{Also in this case we verified that results are basically the same if the resample is done on a time scale of 10 minutes (bias = 0.11~arcsec and RMSE = 0.33~arcsec) with respect to a bias of 0.11~arcsec and a RMSE of 0.36~arcsec for the time scale of 1 minute.}. This figure of merit tells us that the correlation between DIMM and GS is substantially good.
Fig. \ref{fig5}-right shows the cumulative distribution of the same set of measurements in which it is visible that the two data-set have in most cases, the same statistical behavior.
The bias of 0.11 arcsec tells us that the agreement is within 9$\%$ (the mean seeing for the GS and the DIMM are respectively 1.16 and 1.27~arcsec). Looking at Fig. \ref{fig5}-right we observe that for seeing $\le$ 1 arcsec the statistical behavior of the two data-set is almost perfect. A good agreement (within 10$\%$) is observed for the median value of the seeing. A relative error up to 20$\%$ is observed for seeing values in the [1-2]~arcsec range. The relative error is again within 10$\%$ for seeing larger than 2~arcsec. The excess of turbulence measured by the DIMM when the seeing is between 1 and 2 arcsec is due to the fact that the two instruments, even if they are located at the same height above the ground on a substantial flat surface, are separated by a distance of $\sim$ 205~m. The DIMM, in particular, is located at the extremity of the plateau obtained after the cut of the mountain summit performed to build the VLT and it points always at South i.e. towards the summit plateau. It has been noted on routinely observations (Lombardi private communication) that on the ridge of the plateau there is a tendency in creating a local excess of turbulence particularly when the seeing is strong. The local excess relates, of course, turbulence in the surface. Because of the fact that the DIMM points always at South, this excess is typically detected by the DIMM\footnote{It has been observed (Lombardi private communication) that if the DIMM points towards North this effect seems mitigated or even disappears. The GS has not a privileged direction for the lines of sight.}.  This highly probably generates the slightly difference in seeing observed between the GS and the DIMM in the range [1-2] arcsec. We note however that this difference is much weaker in relative terms than that observed between the MASS and the GS. Even more important, the GS/DIMM difference can be justified by the local turbulence in the surface layer that can not be equally sensed by the two instruments but the difference between the MASS and the GS can not be explained by a different line of sight because first, the MASS vs. GS comparison treats just turbulence in the free atmosphere and therefore the argument of the local excess should be a no sense. Second, even if it is well possible that the MASS and the GS are looking at different line of sights in different instants during the night, this fact might eventually cause a temporary (i.e. locally in time) discrepancy but not a systematic bias between the two instruments at specific heights above the ground all along the time. Considering that the GS and the MASS have no privileged lines of sight, if one looks at the problem from a statistical point of view, the potential different line of sight between the two instruments during the night would produce at some instants an excess of turbulence from the GS and in some other instants an excess of turbulence from the MASS and not always an excess of turbulence from the GS. In other words, the line of sight direction is irrelevant to comment the bias observed between the MASS and the GS at some specific heights. With respect to our main analysis goal i.e. the comparison between the MASS and the GS, the cross-comparison between the DIMM and the GS supports the thesis that we are in front of an underestimation of the MASS and not of an overestimation of the GS. The second assumption should imply, indeed, that the GS seeing should be weaker than what is observed with a consequent lack of correlation with the DIMM. 

\begin{table}
\caption{Number of measurements available during each night from the GS and the DIMM to be used for the statistical analysis. \label{table_meas_gsdimm}}
\begin{center}    
\begin{tabular}{c c c }
\hline
Night &  N. of GS meas. & N. of DIMM meas. \\
\hline
10/11/2007 &  322      &   490   \\
11/11/2007 &  361       &  485   \\
13/11/2007 &  290       &  473  \\
14/11/2007 &  356       &  447  \\
17/11/2007 &  303       &  459  \\
18/11/2007 &  391       &  495  \\
20/11/2007 &  371       &  490  \\
21/11/2007 &  147       &   270  \\
22/11/2007 &  320       &   427  \\
24/11/2007 &  361       &   471  \\
15/12/2007 &  299       &   518  \\
16/12/2007 &  423       &   473  \\
17/12/2007 &  345       &   471  \\
18/12/2007 &  454        &  418  \\
19/12/2007 &  440       &   513  \\
20/12/2007 &  483       &   497  \\
21/12/2007 &  492      &    469  \\           
22/12/2007 &  491       &   509  \\
23/12/2007 &  402       &   482  \\
24/12/2007 &  452       &   487  \\
\hline
\end{tabular}
\end{center}
\end{table}

\subsection{Isoplanatic angle: $\theta_{0}$}
\label{mass_gs_iso}

\begin{figure}
\centering
\includegraphics[width=6cm]{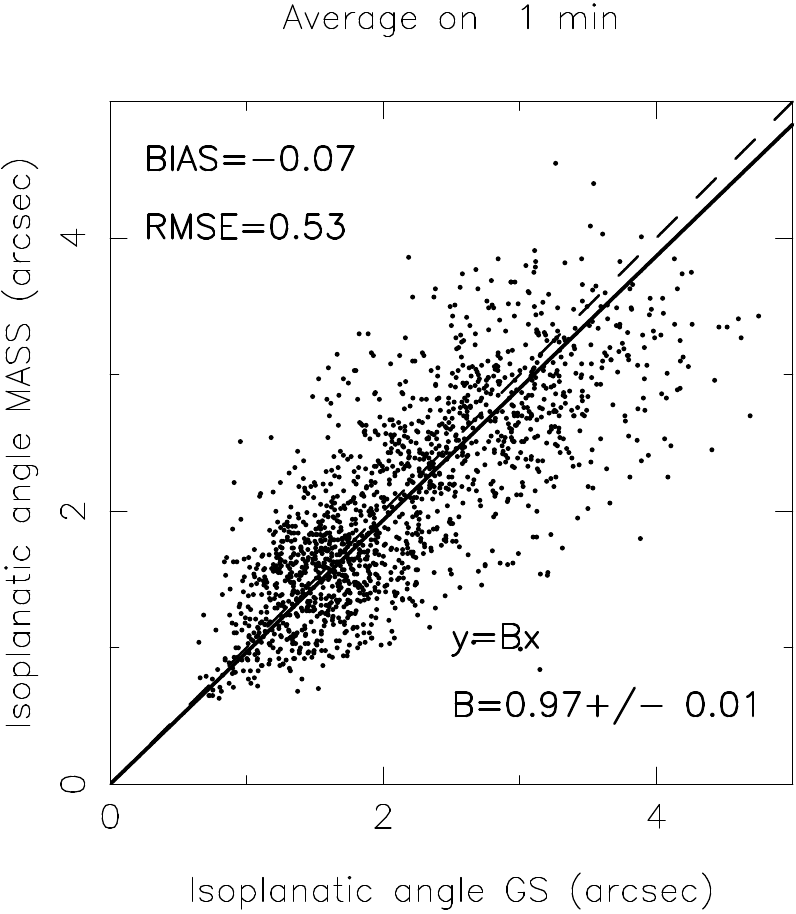}
\caption{Distribution of the MASS and GS measurements of the isoplanatic angle $\theta_0$ related to the nights in which 
simultaneous measurements obtained by the two instruments are available: 12 nights (see Table \ref{tab:nights}). Measurements are 
sampled on a timescale of 1 minute. The bias refers to $\theta_{0,MASS}$ - $\theta_{0,GS}$.
\label{theta0}} 
\end{figure}

Fig. \ref{theta0} shows the isoplanatic angle ($\theta_{0}$) as observed by the MASS and the GS on those nights in which simultaneous observations are available. Measurements are resampled on a time scale of 1 minute and show a bias = -0.07~arcsec and a RMSE = 0.53~arcsec\footnote{We verified that if the resample is done on time scales of 10 and 30 minutes results are very similar for the bias (-0.09~arcsec) and slightly inferior for the RMSE (0.46~arcsec)} with a relative error of -3$\%$ (GS average: $\theta_{0}$ =  2.11~arcsec; MASS average: $\theta_{0}$ =  2.07~arcsec). The $\theta_{0}$ from the GS is obtained through the $\CN2$:\\
\begin{equation}
\theta _{0}=0.057\cdot \lambda ^{6/5}\cdot [\int_{0}^{\infty }h^{5/3}C_{N}^{2}(h)dh]^{-3/5}
\end{equation}
The $\theta_{0}$ from the MASS is retrieved from the calculation of the scintillation indices and it is an output of the software Atmos 2.3. This means that MASS and GS use different methods to calculate $\theta_{0}$.
The sample of the simultaneous measurements between MASS and GS for $\theta_{0}$ is made of 12 nights, the same nights used for the statistical analysis of the seeing. We precise that, because of the method used by the MASS to provide measurements of the seeing and $\theta_{0}$,  an observed seeing value is not automatically associated to the corresponding $\theta_{0}$. This is due to the fact that the $\theta_{0}$ is not obtained by the integral of the turbulence on the whole atmosphere but it is based on the scintillation indices. The simultaneous measurements of seeing and $\theta_{0}$ do not cover therefore necessarily the same temporal range. The value of the bias (Fig.\ref{theta0}) tells us that there is no a systematic error between the two instruments as confirmed also by the regression line. The good agreement observed for the $\theta_{0}$ is not in contradiction with the results obtained for the seeing (Section \ref{mass_gs_see}) that, on the contrary, put in evidence some discrepancies between the two instruments. It is indeed known that $\theta_{0}$ is mainly affected by the turbulence in the high part of the atmosphere because $\theta_{0}$ $\sim$ $h^{5/3}$$\cdot$$\CN2$(h). Our analysis on the seeing told us that the MASS and GS correlation is substantially good in the layer 6 i.e. the highest layer (see Fig. \ref{fig1}, Table \ref{tab:stat} and Fig.\ref{layer1} to Fig.\ref{layer6}). In other words, the OT discrepancies observed in other parts of the atmosphere simply do not affect the $\theta_{0}$ in a significative way from a quantitative point of view. Fig.\ref{iso_te} in Annex shows the temporal evolution of the $\theta_{0}$ in all the 12 nights in which GS and MASS measurements are available. It is possible to observe how the temporal evolution of $\theta_{0}$ as measured by the two instruments is substantially well correlated in terms of measurements trend. We note that, in four of the twelve nights, the statistical sample of measurements is small. It would be, therefore, worth to analyze a richer statistical sample to confirm definitely these conclusions. As a final consideration we remind that, the fact that an agreement is observed from a statistical point of view, does not exclude that on individual nights, even for periods of a few hours, it is possible to have important discrepancies between the two instruments. For example, during the night 20/11 (from around 01:00 and 03:00 UT) and 24/12 (from around 04:00 and 06:00 UT) we observe a relative error of the order of 30-50 $\%$. This is due to the different turbulence estimation in the highest MASS layer (see Fig.\ref{layer6}). This kind of differences, in specific temporal intervals, might be attributed to different lines of sight of the two instruments. It is important to note, however, that, this is just a possible explanation. With that we mean that  this is just an assumption and, in principle, we can not exclude that the difference is not due to a failure of an instrument.
If we assume that the discrepancy is due to different lines of sight of the instruments this would imply the fact that in the high part of the atmosphere (layer 6), there are spatial inhomogeneities in the turbulence distribution on the horizontal plane enough large to produce differences in $\theta_{0}$ of the order of 30-50 $\%$. Such an horizontal inhomogeneity, also in the high part of the atmosphere, would confirm results obtained with a GS by \citet{masciadri2002}\footnote{This was the first evidence in the literature, at least at our knowledge, of the finite horizontal size of turbulence layers at different heights in the 20~km a.g.l. }. It is a fact that the two instruments were not looking at the same direction in that particular interval of time therefore the assumption might be reasonable.

\subsection{Wavefront coherence time: $\tau_{0}$}
\label{mass_gs_tau0}

The wavefront coherence time ($\tau_{0}$) calculated by the GS is obtained from:\newline
\begin{equation}
\tau_{0}=0.057\cdot \lambda^{6/5}\cdot \left [ \int_{0}^{\infty } \left | V(h) \right |^{5/3}\cdot C^{2}_{N}(h)\cdot dh\right ]^{-3/5}
\end{equation}
where the $\CN2$ profiles are measured by the GS and the wind speed vertical profiles have been reconstructed by the atmospherical model Meso-Nh. In previous papers \citep{hagelin2010,masciadri2013a} it has been proved that such an atmospherical model provides very reliable wind speed vertical profiles. The most solid validation of the method has been carried out through a comparison of wind speed profiles provided by the model with radiosoundings on a sample of 53 launches distributed on 23 nights in summer and winter \citet{masciadri2013a}.

\begin{figure}
\centering
\includegraphics[width=6cm]{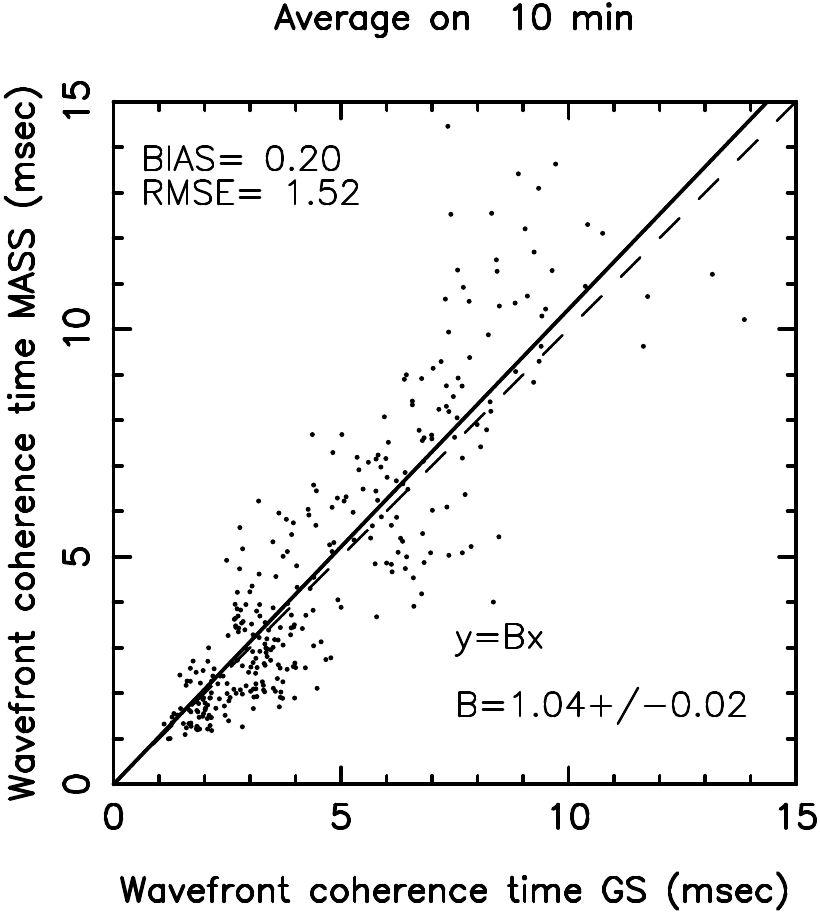}
\caption{Distribution of the MASS and GS measurements of the wavefront coherence time $\tau_0$ related to the nights in which 
simultaneous measurements obtained by the two instruments are available: 14 nights. Measurements are 
sampled on a timescale of 10 minute. The bias refers to $\tau_{0,MASS}$ - $\tau_{0,GS}$.
\label{tau0}} 
\end{figure}

The MASS wavefront coherence time on the whole atmosphere (hereafter total MASS $\tau_{0,tot}$) is defined as \citep{kornilov2011}:
\begin{equation}
\tau_{0,tot}=\left [ (\tau_{0,GL})^{-5/3}+(\tau_{0,MASS})^{-5/3} \right ]^{-3/5}
\label{eq:tau_mass}
\end{equation}
where $\tau_{0,MASS}$ comes from the MASS covering the contribution of the free atmosphere and:\newline
\begin{equation}
\tau_{0,GL}=0.314\cdot \frac{r_{0,GL}}{V_{30m}}
\end{equation}
is the $\tau_{0}$ contribution associated to the boundary layer. $r_{0,GL}$ is the Fried parameter associated to the near-ground contribution of r$_{0}$ obtained subtracting the free atmosphere turbulent energy (MASS) from the total turbulent energy (DIMM). $V_{30m}$ is the wind speed measured at 30~m above the ground by the Automatic Weather Station (AWS) \citep{sandrock99}. $r_{0,GL}$ is obtained from:\newline
\begin{equation}
r_{0,GL}=\left [ (r_{0,DIMM})^{-5/3}-(r_{0,MASS})^{-5/3} \right ]^{-3/5}
\end{equation}
\noindent
For the estimation of the MASS $\tau_{0,MASS}$ we used the Atmos 2.97.3 software version that implements a different method for the calculation of $\tau_{0}$ with respect to what was done in the Atmos 2.3 version \citep{kornilov2011}\footnote{The new $\tau_{0}$ is obtained indeed, taking into account a finite exposure time and a weighting function depending on the wind speed. }. It is known that the Atmos 2.3 version used for the seeing and $\theta_{0}$ does not provide reliable $\tau_{0}$ estimates \citep{kornilov2011}. For the $\tau_{0,tot}$ we have simultaneous measurements obtained with the MASS and the GS on a sample of 14 nights (all the 12 nights included in the sample used for the seeing and $\theta_{0}$ (Table \ref{tab:nights}) plus the nights of 15/12/2007 and 16/12/2007. Fig. \ref{tau0} shows the scattering plot of the $\tau_{0}$ as observed by the MASS and the GS on those nights in which simultaneous observations are available. Measurements are resampled on a time scale of 10 minutes\footnote{Due to the fact that the MASS $\tau_{0,tot}$ is obtained using measurements coming from the DIMM, the MASS and the AWS, we resampled the measurements on a time scale that is a common multiple of the sampling of the individual data-set.}. We can observe that the agreement between the two instruments is substantially good with a bias = 0.20~msec with a relative error of 1$\%$ (GS average: $\tau_{0}$ = 4.49~msec; MASS average: 4.54~msec). Fig.\ref{tau0_te} in Annex shows the temporal evolution of the $\tau_{0}$ in all the 14 nights in which GS and MASS measurements are available. It is possible to observe how the temporal evolution of $\tau_{0}$ as measured by the two instruments is substantially well correlated also in terms of measurements trend. If we look at the individual nights we note in a couple of nights (23/12 and 24/12) an important relative error (even $>$ 60 $\%$) in the interval of a few hours.

However, because of the fact that the $\tau_{0}$ comes from the product of the wind speed and the $\CN2$ profiles and considering that we observed some discrepancies of the turbulence ($\CN2$) detected on individual layers and on the free atmosphere by the MASS and the GS, one could wonder how an agreement of the MASS and GS $\tau_{0}$ is possible. One should expect an overestimation of the $\tau_{0}$ from the MASS because the MASS underestimate the seeing in the free atmosphere. Why is this not observed ? A probable explanation is the following. $\tau_{0,tot}$ is given by the addition of two contributions: the first one comes from the MASS ($\tau_{0,MASS}$); the second one comes substantially from the DIMM ($\tau_{0,GL}$) ($r_{0,MASS}$ is indeed much weaker than $r_{0,DIMM}$). Looking at Fig. \ref{tau0_te}, we note that, in the second part of the campaign (starting from 19/12), where the absolute value of $\tau_{0}$ is large, the MASS $\tau_{0}$ shows some tendency in giving a larger $\tau_{0}$ value with respect to the GS. We verified (Fig. \ref{tau0_gl_vs_fa}) that the nights characterized by a small $\tau_{0}$ ($\tau_{0}$ $<$ 5 msec) correspond to the cases in which the $\tau_{0,GL}$ contribution is dominant with respect to the $\tau_{0,MASS}$. The term 'dominant' means smaller because the $\tau_{0}$ respects Eq.\ref{eq:tau_mass}.
In the nights characterized by a large $\tau_{0}$ ($\tau_{0}$ $>$ 5 msec) we can have both cases even if it is more frequent that the $\tau_{0,MASS}$ contribution is dominant. This indicates that the turbulence and/or the wind speed (or even both of them) in the boundary layer are so weak that the $\tau_{0}$ contribution from the free atmosphere becomes dominant. In those cases the overestimation of the MASS starts to be evident but on the whole sample of nights the DIMM contribution is dominant with respect to the MASS one. Considering that the typical $\tau_{0}$ median value of a good site is of the order of 4-5 msec, i.e. close to the threshold that we indicated, we think that it would be very useful, for precautionary principle,  to do in the future the same comparison of MASS vs. GS $\tau_{0}$ on a richer statistical sample to confirm the conclusion of a good correlation and to confirm the fact that the overestimation tendency of the MASS remains statistically not so important.

\section{Discussion}
\label{disc}

We observe that, even if our results as well as those obtained by \citep{tokovinin2005} (hereafter T05) show some discrepancies between the MASS and the GS measurements, the two studies reached in some cases similar results, in other cases different results. Table \ref{tab:stat} columns 8 and 9 report the relative errors calculated in this paper and in the T05 paper with respect to the integrated turbulence J in the free atmosphere and in the individual layers. As already said in Section \ref{mass_gs_see} the relative errors expressed in seeing (instead of J) are smaller (see col.7 of Table \ref{tab:stat}) but we calculated also the relative errors in J because T05 used J. A premise is necessary: due to the fact that the statistical samples of the two studies are relatively small and in any case not equal (12-14 nights for our study and 4 nights for the T05 study) we can not expect necessarily equivalence in the absolute value of the relative errors. What should be reasonable to expect is, however, at least a similar behavior for example, the evidence of an excess or a default or a matching of turbulence in specific regions of the atmosphere. 

For what concerns the vertical distribution of J in the individual layers the similarities between the two studies are: {\bf (a)} in the layer located at 16~km (layer 6) we both found a good agreement with a relative error within +20$\%$ (this paper) and within -25$\%$ (T05 paper) even if in opposite direction; {\bf (b)} in the layer located at 1~km (layer 2) we both found that MASS underestimates the seeing with respect to the GS with a relative error of -71$\%$ (this paper) and -41$\%$ (T05 paper); {\bf (c)} we both found a substantial underestimation of J by the MASS with respect to the GS in the layer located at 4~km (layer 4) with a relative error of -60$\%$ (this paper) and -57$\%$ (T05).\\
Differences between the two studies are: {\bf (a)} we found that the MASS underestimates J in the layer located at 0.5~km (layer 1) with a relative error of -82$\%$ while \citet{tokovinin2005} found that the MASS overestimates the GS with a relative error of +100$\%$; {\bf (b)} we found a substantial good agreement of J in layer located at 2~km (layer 3) with a relative error within +25$\%$ while T05 found that the MASS underestimates J with respect to the GS with a relative error of -67$\%$; {\bf (c)} we found that the MASS underestimates J in the layer located at 8~km (layer 5) with a relative error of -71$\%$ while T05 found a good agreement between the two instruments with a relative error within +14$\%$; {\bf (d)} we found a substantial underestimation by the MASS with respect to the GS of the J in the free atmosphere seeing (obtained as the contribution coming from all the six layers) with a relative error of -48$\%$ while T05 found a substantial good agreement between the two instruments with a relative error of -5$\%$. 

In synthesis our study put in evidence the fact that in almost all the layers (with exception of layers located at 2 and 16~km i.e. layers 3 and 6) there are important discrepancies between the turbulence observed by the GS and the MASS. Also we observe that the integration of the turbulence in the free atmosphere is underestimated by the MASS by a not negligible factor (-48$\%$ in J terms and -32$\%$ in seeing terms). T05 founded some not negligible discrepancies in layers located at 0.5 and 4~km but a substantial good agreement in estimating the free atmosphere by the two instruments (within 5$\%$).

The most important difference between the results of the two studies is, therefore, the result on the free-atmosphere integration. The sample of 4 nights of T05 provides a statistical substantial good agreement while we obtain a statistical substantial underestimation of the MASS with respect to the GS on a sample of 12 nights. Our results are, in principle, more reliable from a statistical point of view (12 nights against 4 nights). We think, however, that it would be useful to repeat the calculation on a richer statistical sample to confirm these results. They indicate, indeed, biases on the MASS estimations when it is used as a monitor of the optical turbulence in the free-atmosphere. It is therefore important to fix this point to use correctly this instrument in this context. 

The other differences found in the two studies are more difficult to be justified. We think however that the most worried thing is the fact that visibly important discrepancies between the layers are detected in both studies. A richer statistical sample would certainly help in better understanding the status of art. We are promoting, in the context of an international working group \citep{masciadri2013b}, a site testing campaign aiming at establish absolute calibration of different vertical profilers among those, the MASS and the GS. It is evident that such not negligible discrepancies in individual layers can not be considered satisfactory for some applications such as the one in which we are interested on, for example the identification of a reference for a model validation. Discrepancies on individual layers are also very critical in application to the optimization of all kind of AO systems that are particularly sensitive to the position and strength of the individual turbulent layers. For the same reasons the MASS appears not very suitable for the calibration of the atmospherical models for the OT forecast.

We report here few words on the consequences of our results on previous studies done in the context of the site search/selection. We note that, in spite of evident OT discrepancies on individual layers, $\theta_{0}$ appears well correlated between the MASS and the GS. There are therefore no major implications that might contradict results obtained previously. The free atmosphere seeing was probably underestimated (and the boundary layer overestimated because it is calculated by subtraction of the MASS contribution from the DIMM one) but we note that this happened for all the sites and, due to the fact that the site selection is an exercise done in relative terms, the consequences are therefore not so critical. For the $\tau_{0}$ we remind that, at the epoch of the TMT and E-ELT site selection the Atmos 2.97.3 software that we used was not available. The two teams used different approaches to calculate $\tau_{0}$. In the TMT case \citep{travouillon2009} the MASS software was corrected by a constant factor (1.73) found fitting measurements with wind speed radiosoundings and the order of magnitude of the final $\tau_{0}$ was included between 4.2 and 5.6 msec. In the E-ELT case \citep{vazquez-ramio2012} $\tau_{0}$ was calculated using the $\CN2$ from MASS and DIMM plus the wind speed from the Global Data Assimilation System (GDAS) database, with a 1$^{\circ}$ horizontal resolution of the Air Resources Laboratory (ARL) of the National Oceanic and Atmospheric. It is known, however, that these data coming form the General Circulation Models have a too low resolution to well represent the wind speed in the low part of the atmosphere, particularly on mountain regions \citep{masciadri2013a}. Limiting our commentaries to the potential consequences of the results of this paper on previous $\tau_{0}$ estimations we can simply say that, apart the fact that in both cases they used the $\CN2$ in the free atmosphere from the MASS that was, highly probably, underestimated, it would be a hazard to deduce any other consequences from our results. We limit ourselves to observe that the order of magnitude of the $\tau_{0}$ median values found in those studies fit with the expected order of magnitude. 

\section{Conclusions}
\label{concl}

In this study we present the results of a detailed comparison of the turbulence stratification in the free atmosphere as well as of the astroclimatic parameters (seeing, isoplanatic angle and wavefront coherence time) as measured by the MASS and the GS. The main results that we obtained tell us that:\newline

\begin{itemize}
\item the MASS underestimates the integrated turbulence (J or seeing) in the free atmosphere with respect to the GS with a relative error of -32$\%$ in seeing terms (-48$\%$ in J terms); a bias = -0.21~arcsec and a RMSE = 0.25~arcsec if we consider the 6 layers of the MASS. If we consider the highest 5 layers (from layer 2 to layer 6) the relative error decreases to -20$\%$ in seeing terms (-30$\%$ in J terms); the bias = -0.1 arcsec and the RMSE = 0.16~arcsec. We find important discrepancies between MASS and GS in all the individual layers (reaching relative errors as high as -65$\%$ in seeing terms and -82$\%$ in J terms) with exception of the layers located at 2 and 16~km (layers 3 and 6) in which the relative error remains limited to +18$\%$ in seeing terms (+20$\%$ in J terms);
\item the unique study previously published on a similar topic (even if it was applied on a poorer statistical sample, \citet{tokovinin2005}) revealed relative errors on individual layers as large as those we observed in our study;
\item the main difference between our results and those of \citet{tokovinin2005} is that they found a substantial good agreement of the free-atmosphere seeing as measured by the MASS and the GS on a statistical sample of 4 nights while we find that the MASS underestimates the free-atmosphere seeing with respect to the GS on a statistical sample of 12 nights. The relative error is -32$\%$ in seeing terms (-48$\%$ in J terms). Even if our estimations are statistically more reliable, it should be useful to confirm our results on a richer statistical sample. This result, if confirmed, should have important consequences on the main use of the MASS done so far i.e. as a monitor of the free-atmosphere seeing;
\item the isoplanatic angle ($\theta_{0}$) appears substantially well correlated between the two instruments (bias = -0.07~arcsec; RMSE = 0.53~arcsec and relative error r = 3$\%$) because the most important contribution to $\theta_{0}$ comes from the OT in the highest layer (layer 6) that is well correlated between the two instruments;
\item the wavefront coherence time ($\tau_{0}$) is well correlated between the two instruments (bias = 0.20~msec; RMSE = 1.52~msec and relative error r = 1$\%$). The underestimation of the MASS in quantifying the free atmosphere turbulence seems to have a minor effect on the $\tau_{0}$ because the boundary layer contribution from the DIMM ($\tau_{0,GL}$) has, in many cases, a dominant role with respect to the contribution from the MASS ($\tau_{0,MASS}$). It should be useful to repeat the same calculation on a richer statistical sample to confirm this thesis. We observed that, the cases in which biases induced by the overestimation of the free atmosphere turbulence from the MASS can be more evident, is when $\tau_{0}$ is large ($>$ 5 msec);
\item we proved that, contrary to what is frequently believed, the boundary layer contribution of the $\tau_{0}$ has, most of the time, a more important impact from a quantitative point of view than the free atmosphere one.;
\item even if the good correlation of some astroclimatic parameters has been observed in statistical terms ($\theta_{0}$ and $\tau_{0}$) this does not exclude that, on individual nights, on intervals of a few hours, important discrepancies are observed between the two instruments. We therefore warn in using appropriately the conclusions of this study depending on the contexts;  
\item we proved that conclusions reached in this study definitely would require a richer statistical sample of measurements to be confirmed. An international action is on-going \citep{masciadri2013b} to promote a site testing campaign including classical and new-generation vertical profilers aiming at quantifying the absolute values of the optical turbulence and the intrinsic uncertainty we have to consider for each astroclimatic parameter and at validating the new-generation instruments. The intention is to collect measurements on rich statistical samples (order of two weeks in summer and two weeks in winter). The MASS is supposed to be one of these instruments i.e. we will be able to verify these results on a richer statistical sample. Besides, such great relative errors on individual layers as well as on the free atmosphere suggests ancillary studies based on dedicated simulations to identify the origin of the MASS discrepancies;
\item our results indicate that the MASS is not necessarily a suitable tool for some applications such as the calibration of atmospherical model for the OT forecasts and all applications in which it is crucial to know the location and the OT strength of individual layers such as the AO;
\item concerning the impact that these results can have on site selection studies previously done (TMT and E-ELT) we showed that in some cases there were no consequence at all. In other cases we suspect some biases but, due to the fact that the site selection is a relative terms exercise and therefore the biases should be applied to all sites, we think that these results should have no major consequences on the most important goal of these activities i.e. the selection of the site for the respective telescopes.
\end{itemize}

\section*{Acknowledgments}
This study is co-funded by the ESO contract: E-SOW-ESO-245-0933 (MOSE Project).
We are very grateful to the ESO Board of MOSE (Marc Sarazin, Pierre-Yves Madec, Florian Kerber and Harald Kuntschner) for their constant support to this study. 
This study takes advantage of measurements performed during the PAR2007 site testing campaign promoted by ESO in the context of the E-ELT Design Study (FP6 Program). Different international teams from Laboratoire Lagrange - Universit\'e de Nice-Sophia Antipolis (France), IAC (Spain), ESO (Germany), Sternberg Astronomical Institute (Russia) participated and/or have been involved in the PAR2007 campaign.

\appendix

\section{Temporal evolution of astroclimatic parameters}

In this Section the temporal evolution of J (integrated on the free atmosphere and on individual layers defined by the MASS weighting funcitons), isoplanatic angle $\theta_{0}$ and wavefront coherence time $\tau_{0}$ measured by the MASS and the GS are reported for all those nights for which MASS and GS measurements are available in the  PAR2007 Site Testing campaign.

\begin{figure*}
\begin{center}
\includegraphics[width=15cm]{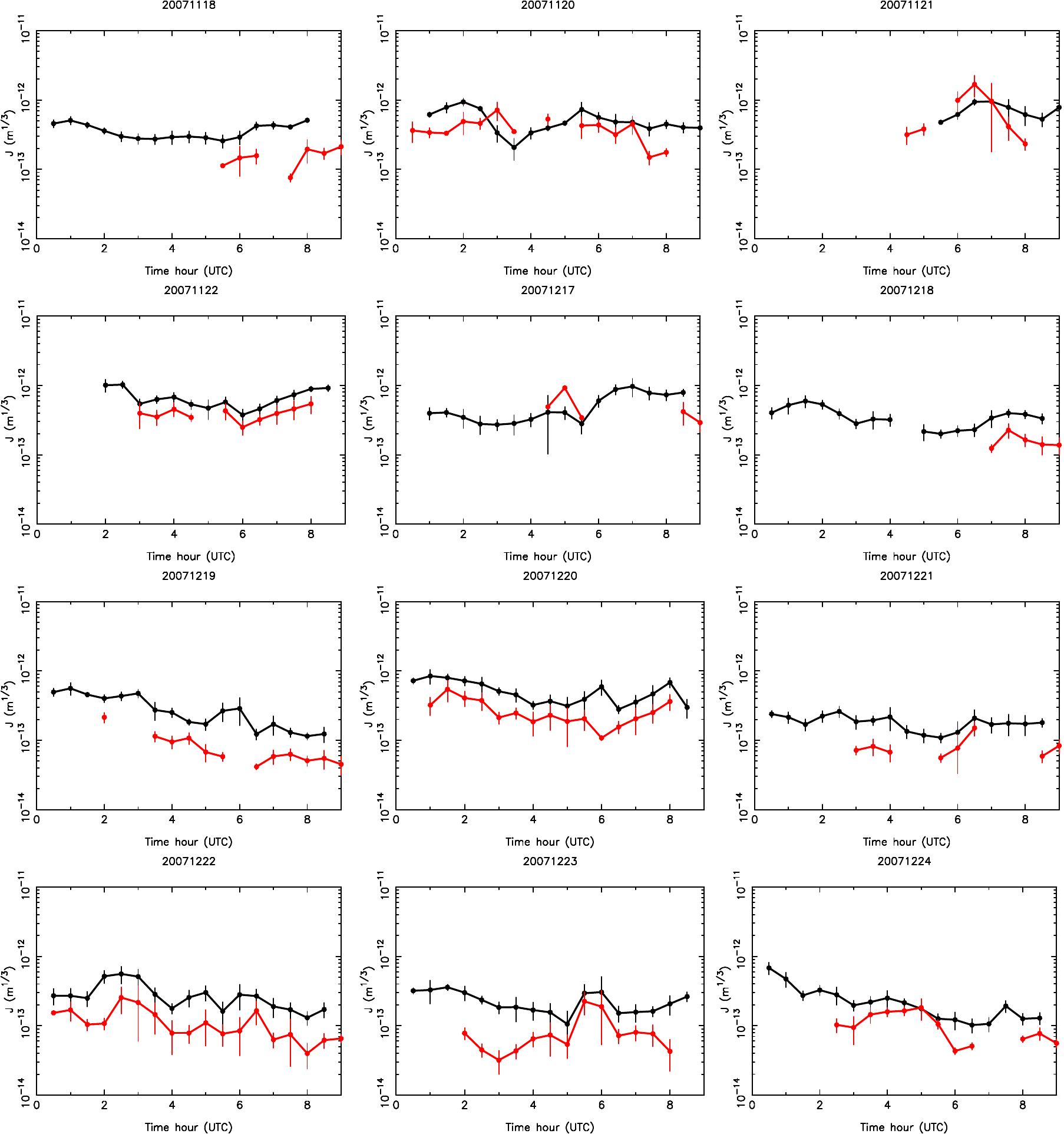}
\end{center}
\caption{Temporal evolution of the GS (black line) and MASS (red line) measurements of the total turbulence (J values) during the 12 nights in which
measurements from both instruments are available. Contributions from all the 6 layers are reported. Measurements are averaged on 30 minutes. Vertical bars are the standard deviation: $\pm$ $\sigma$.
\label{fig4}} 
\end{figure*}

\begin{figure*}
\begin{center}
\includegraphics[width=14cm]{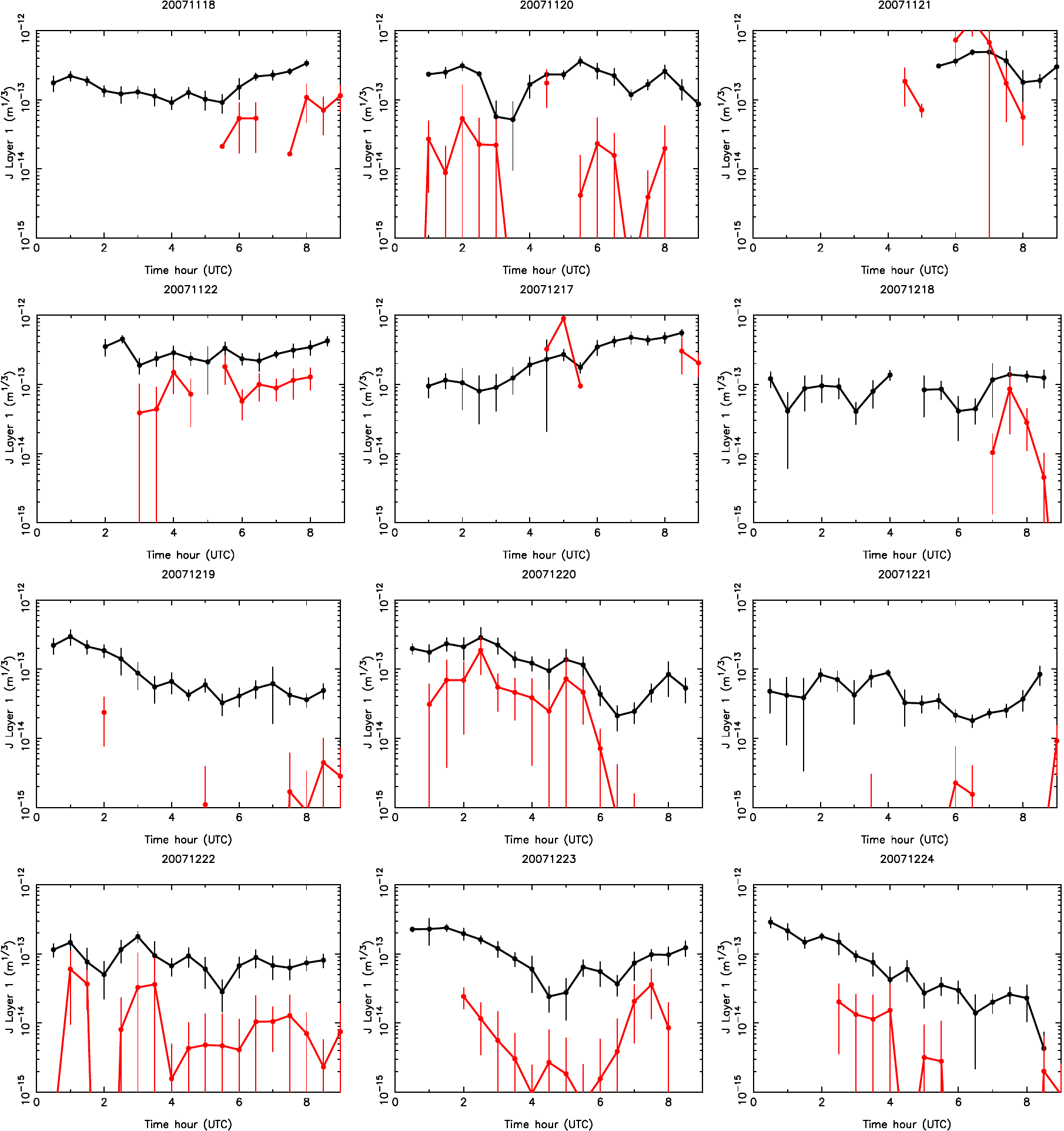}
\end{center}
\caption{Temporal evolution of the GS (black line) and MASS (red line) measurements (J values) in the layer located at 0.5~km (layer 1) during the 12 nights in which
measurements from both instruments are available. Measurements are averaged on 30 minutes. Vertical bars are the standard deviation: $\pm$ $\sigma$.
\label{layer1}} 
\end{figure*}

\begin{figure*}
\begin{center}
\includegraphics[width=15cm]{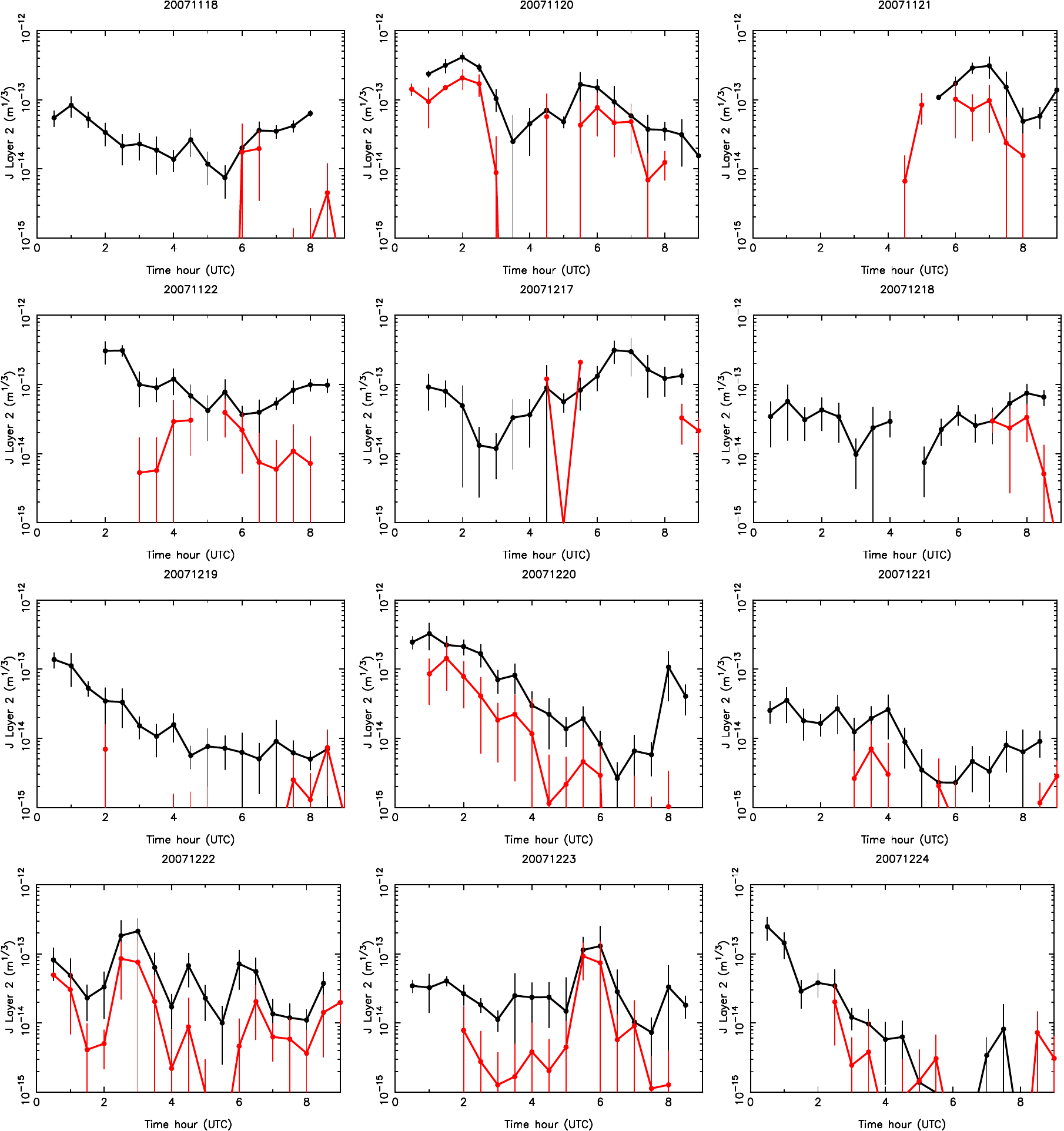}
\end{center}
\caption{As Fig.\ref{layer1} but for the layer located at 1~km (layer 2).
\label{layer2}} 
\end{figure*}

\begin{figure*}
\begin{center}
\includegraphics[width=15cm]{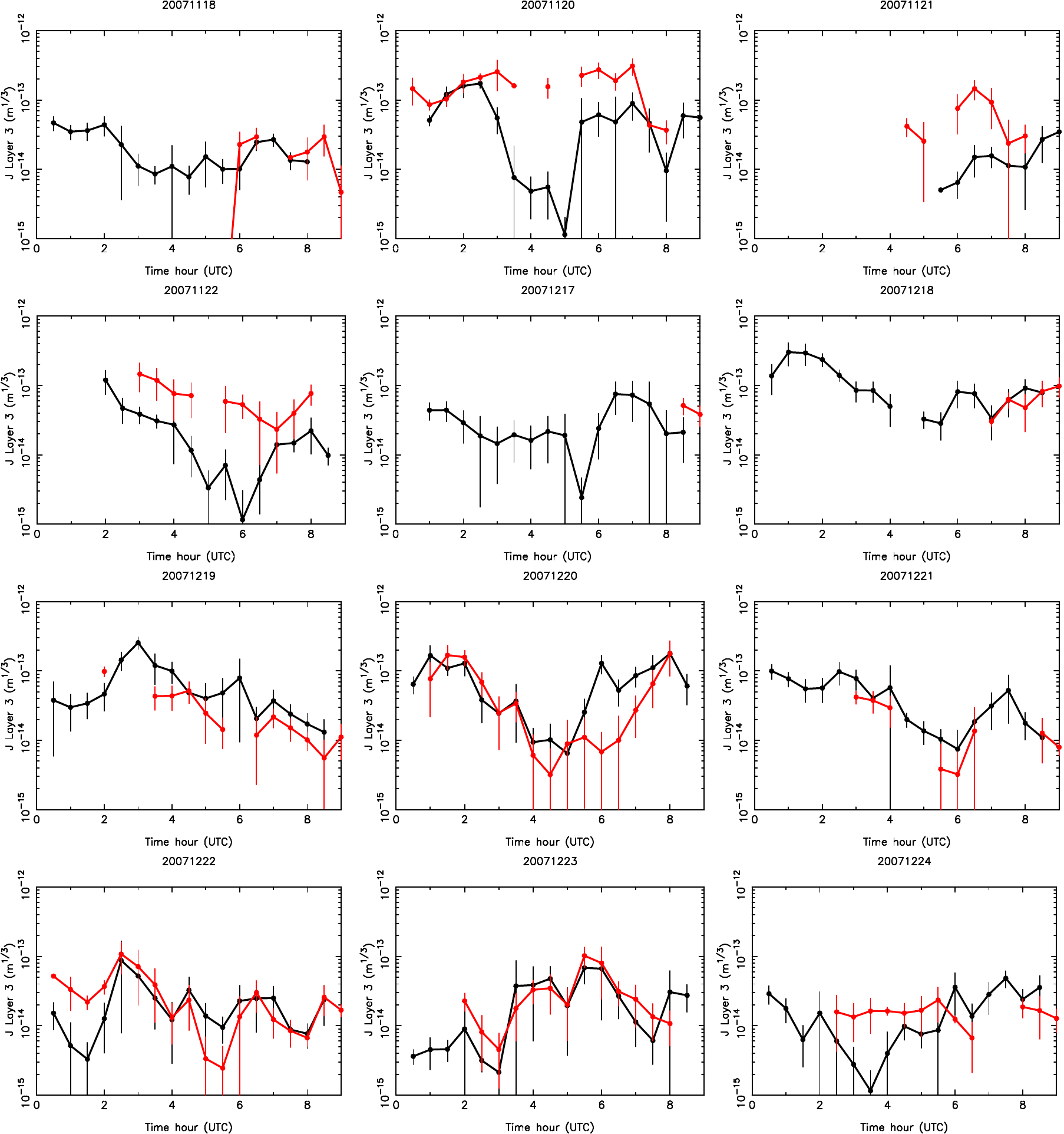}
\end{center}
\caption{As Fig.\ref{layer1} but for the layer located at 2~km (layer 3).
\label{layer3}} 
\end{figure*}

\begin{figure*}
\begin{center}
\includegraphics[width=15cm]{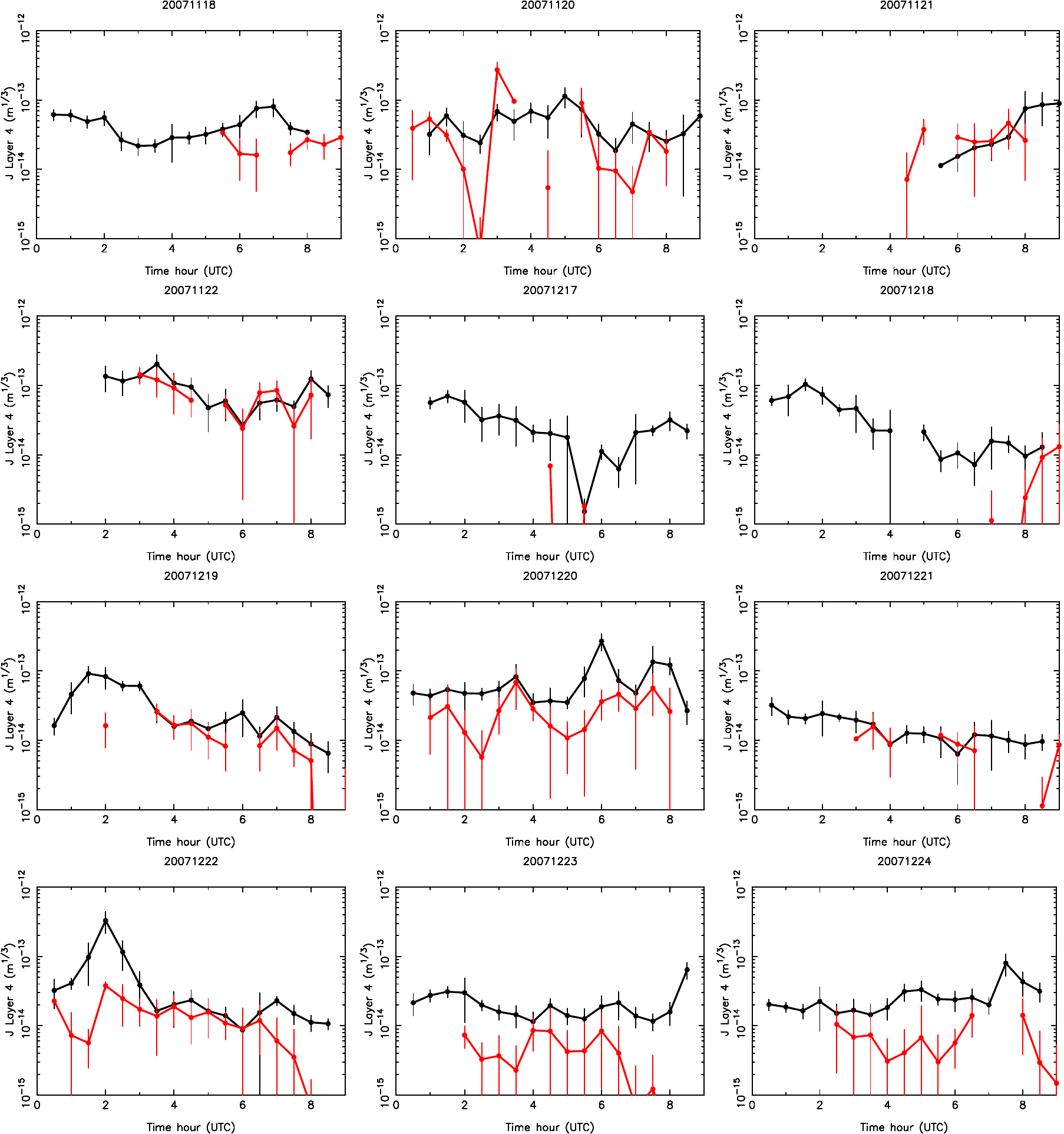}
\end{center}
\caption{As Fig.\ref{layer1} but for the layer located at 4~km (layer 4).
\label{layer4}} 
\end{figure*}

\begin{figure*}
\begin{center}
\includegraphics[width=15cm]{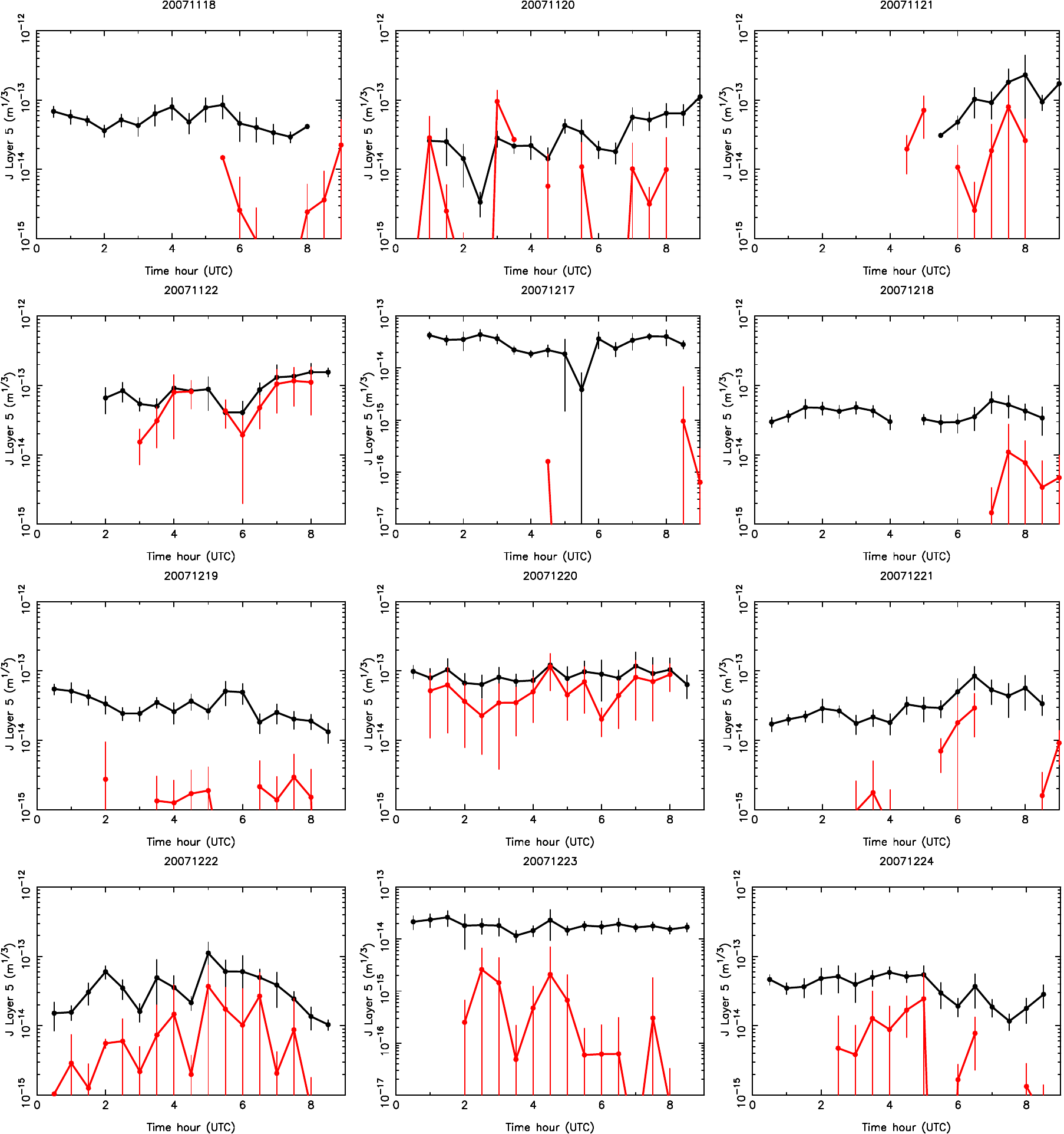}
\end{center}
\caption{As Fig.\ref{layer1} but for the layer located at 8~km (layer 5). {\bf Warning:} nights 17/12/2007 and 23/12/2007 are displayed on the y-axis on the [10$^{-17}$,10$^{-12}$] range.
\label{layer5}} 
\end{figure*}

\begin{figure*}
\begin{center}
\includegraphics[width=15cm]{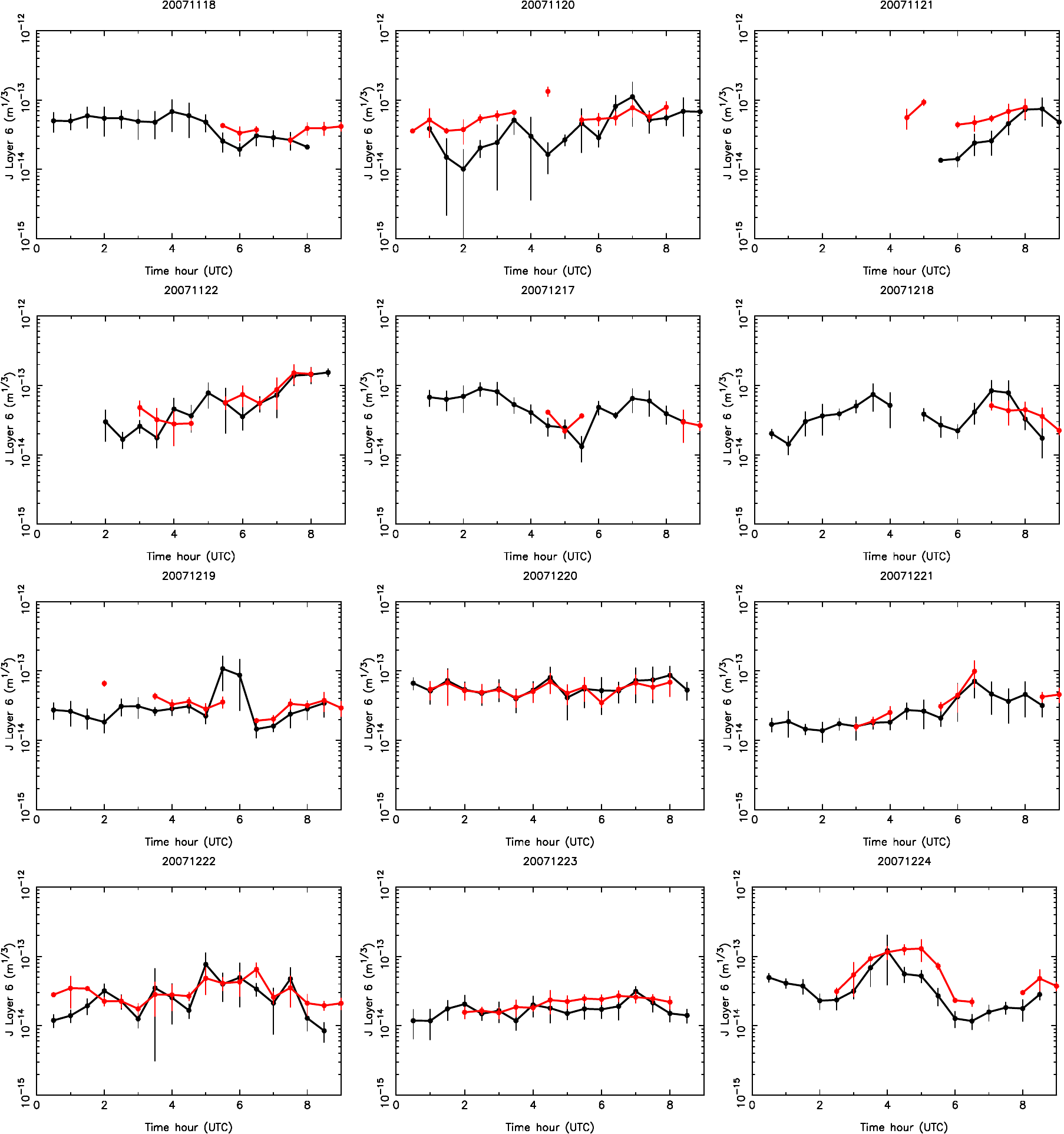}
\end{center}
\caption{As Fig.\ref{layer1} but for the layer located at 16~km (layer 6).
\label{layer6}} 
\end{figure*}

\begin{figure*}
\begin{center}
\includegraphics[width=15cm]{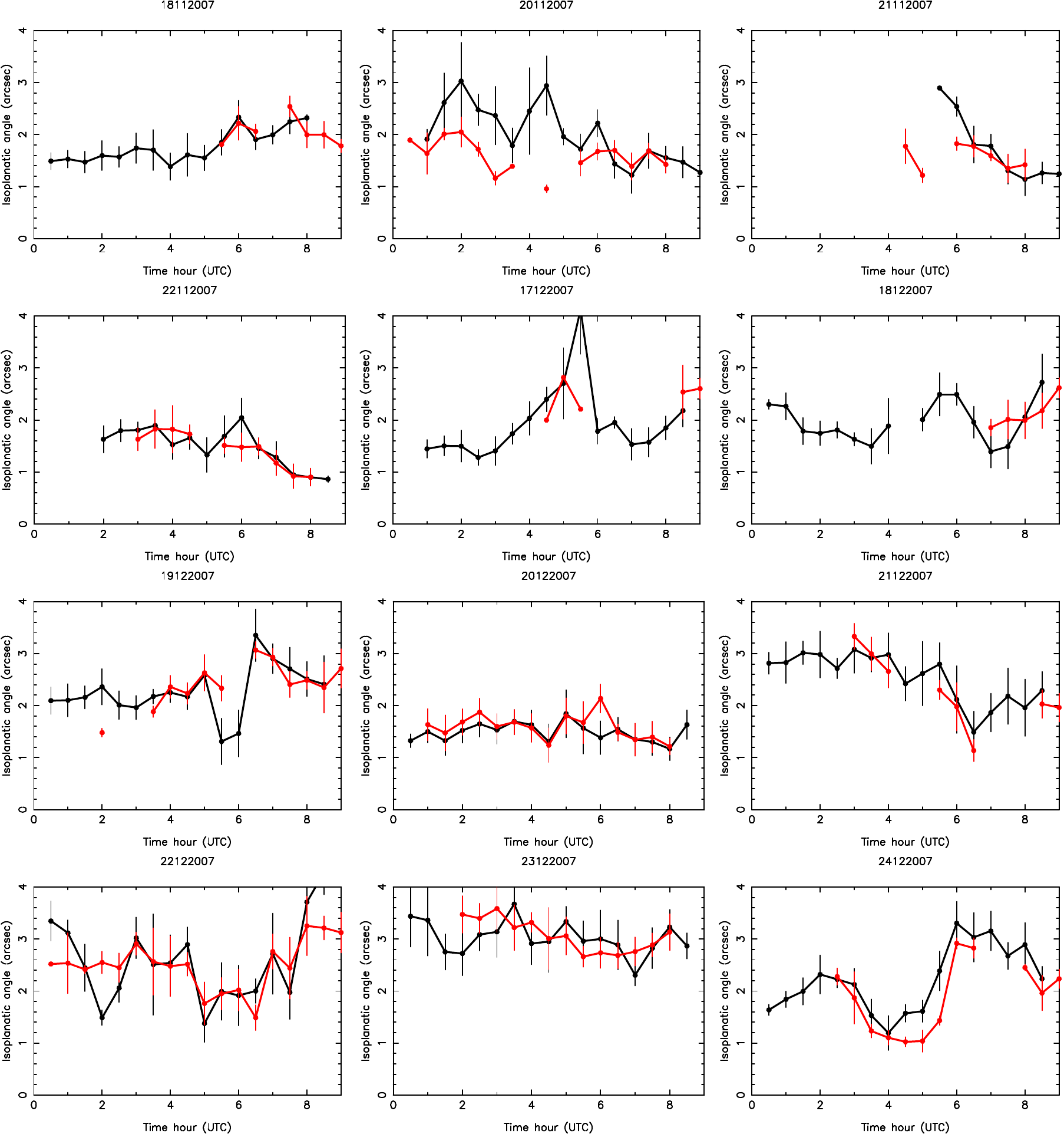}
\end{center}
\caption{Temporal evolution of the GS (black line) and MASS (red line) isoplanatic angle ($\theta_{0}$) during the 12 nights in which
measurements from both instruments are available. Measurements are averaged on 30 minutes. Vertical bars are the standard deviation: $\pm$ $\sigma$.
\label{iso_te}} 
\end{figure*}

\begin{figure*}
\begin{center}
\includegraphics[width=15cm]{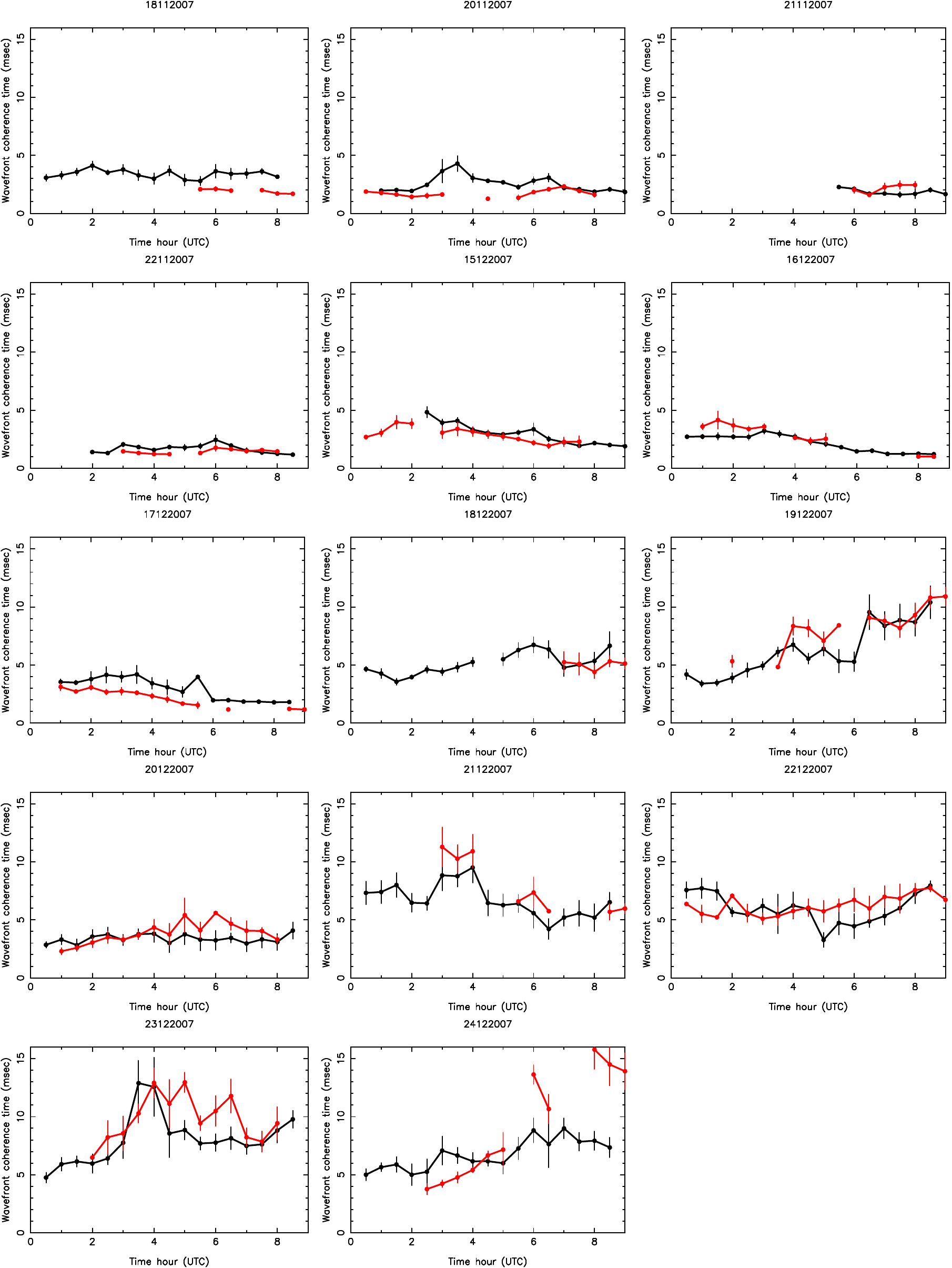}
\end{center}
\caption{Temporal evolution of the GS (black line) and MASS (red line) wavefront coherence time ($\tau_{0}$) during the 14 nights in which
measurements from both instruments are available. Measurements are averaged on 30 minutes. Vertical bars are the standard deviation: $\pm$ $\sigma$.
\label{tau0_te}} 
\end{figure*}

\begin{figure*}
\begin{center}
\includegraphics[width=15cm]{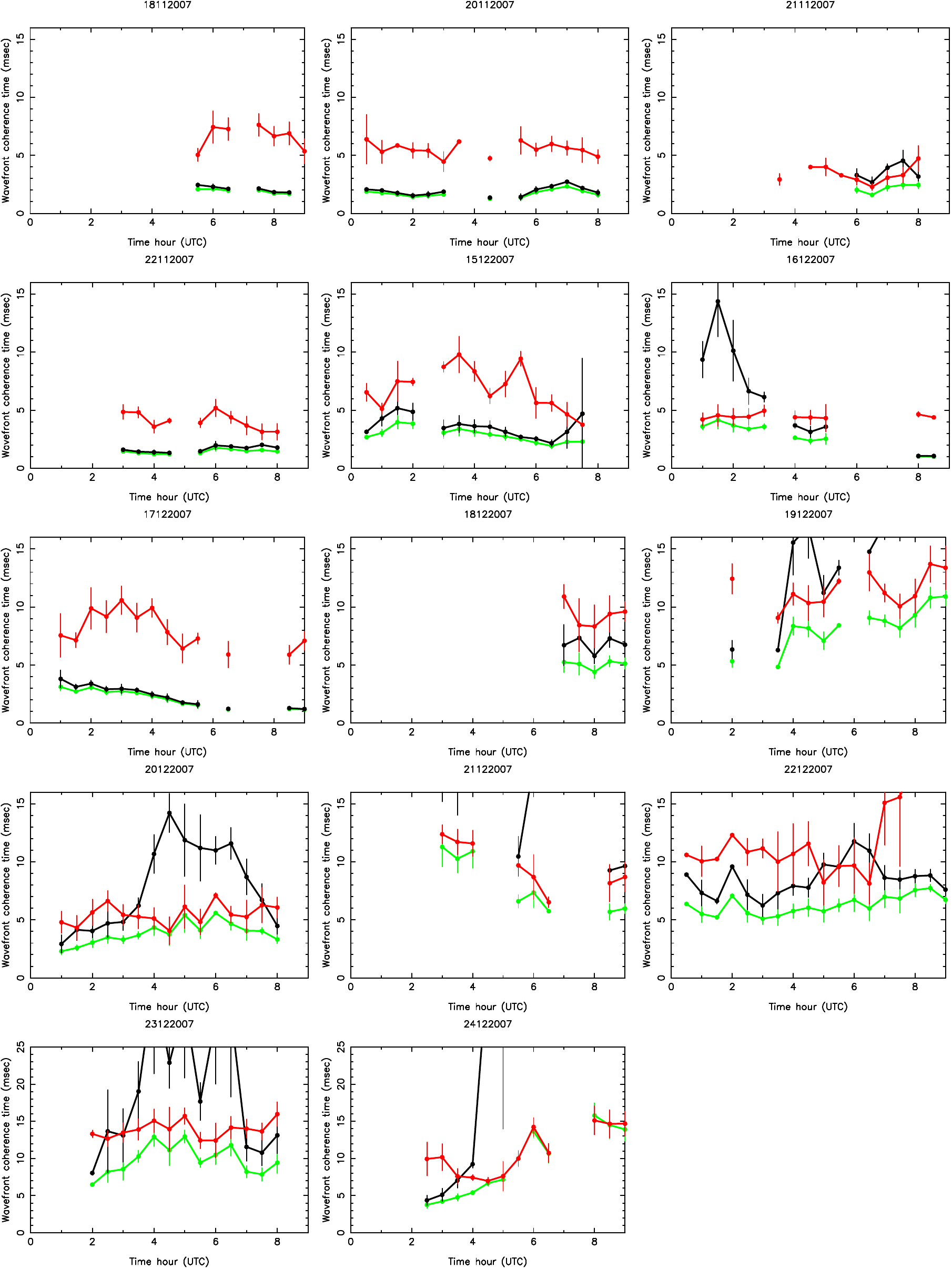}
\end{center}
\caption{Temporal evolution of the $\tau_{0,GL}$ (black line), $\tau_{0,MASS}$ (red line) and $\tau_{0,tot}$ (green line) during the 14 nights in which
measurements from both instruments are available. Measurements are averaged on 30 minutes. Vertical bars are the standard deviation: $\pm$ $\sigma$. {\bf Warning:} nights 23/12/2007 and 24/12/2007 are displayed on the y-axis on the [0,25]~ms range.
\label{tau0_gl_vs_fa}} 
\end{figure*}

\end{document}